\documentclass[aps, prr, amsmath,amssymb,aps,floatfix, twocolumn, superscriptaddress]{revtex4-2}

\usepackage[utf8]{inputenc}
\usepackage{graphicx}
\usepackage{dcolumn}
\usepackage{bm}
\usepackage{color}
\usepackage[normalem]{ulem}

\def\spose#1{\hbox to 0pt{#1\hss}}
\def\lesssim{\mathrel{\spose{\lower 3pt\hbox{$\mathchar"218$}}
 \raise 2.0pt\hbox{$\mathchar"13C$}}}
\def\gtrsim{\mathrel{\spose{\lower 3pt\hbox{$\mathchar"218$}}
 \raise 2.0pt\hbox{$\mathchar"13E$}}}
\definecolor{darkgreen}{rgb}{0.0, 0.5, 0.0}

\begin{document}

\title{Recurrence in Sequence Space: How Epistasis Mimics Shared Ancestry}

\author{Pietro Chiarantoni}
\author{Mina Mahboubi}
\author{Jason Lamanna}
\author{Vincenzo Carnevale}
\affiliation{Institute for Computational Molecular Science, Temple University, Philadelphia, PA, 19122}
\affiliation{Institute for Genomics and Evolutionary Medicine, Temple University, Philadelphia, PA, 19122}
\affiliation{Department of Biology, Temple University, Philadelphia, PA, 19122}

\date{\today}

\begin{abstract}

Sequences in a protein family do not explore the sequence space freely: folding, packing, and binding couple residues, so the set of viable sequences is irreducibly correlated, a phenomenon referred to as epistasis. We show that such correlations lower the \emph{intrinsic dimension} $d_{\rm IN}$ of sequence space, counteracting the concentration of measure and broadening the distribution of pairwise distances. Consequentially, they make it likely that independently evolving lineages \emph{recur}, approaching one another by chance long after they last shared an ancestor. Using Kac's lemma we show that the logarithmic mean recurrence time grows linearly with the intrinsic dimension, $\log \tau_{\rm rec} \propto {d_{\rm IN}}$ for typical sequences, so strongly constrained (low-dimensional) families recur often and generate sequence similarity that carries no genealogical signal. We establish this first in a solvable Ising-like model with tunable correlation, then in a real protein family using Potts models and Markov chain Monte Carlo, where $d_{\rm IN}$ remains predictive of recurrence even in multi-basins landscapes. Because recurrence is a property of the evolutionary process rather than of any estimator, it bounds the recoverability of deep history for distance-based and likelihood methods alike.

\end{abstract}

\maketitle

\section{Introduction}

A recurring theme in molecular evolution is the tension between history and constraint~\cite{Jacob1977,MaynardSmith1970,Harms2013,Blount2018}. History is the result of contingent branching: mutation, drift, duplication, and speciation continually generate variants of a gene, scattering related sequences across genomes and across the tree of life~\cite{Kimura1983,Ohno1970,Fitch1970,Koonin2005}. Constraint reflects the requirement to remain functional: a protein must fold with sufficient stability, pack its core, bind its partners, and, for many families, gate, signal, and allosterically couple distant sites. These requirements sharply restrict which mutations are tolerated and, ultimately, which are fixed. The diversity displayed by a protein family is therefore never a free exploration of sequence space but a walk confined to the narrow, irregularly shaped subset of sequences that satisfy the constraints.

Crucially, these constraints are collective. Stability, packing, and binding are properties of the whole molecule, so the effect of a substitution at one site depends on the residues present at many others. Mutational effects are thus contingent on genetic background, and evolution becomes intrinsically context-dependent, a phenomenon referred to as epistasis~\cite{Starr2016,Breen2012,deVisser2014,Pollock2012,Shah2015,Patel2022}. Concretely, epistasis means that the set of viable sequences is not a Cartesian product of single-site alphabets: knowing which residues are individually admissible at each position is not enough to determine which combinations are admissible jointly. The viable ensemble is in this sense irreducibly correlated: it cannot be reconstructed from any collection of single-site statistics, and it is these correlations, not single-site conservation, that a set of homologs actually samples. These correlations can give rise to hierarchy. Consider a single position at which two amino acids are admissible: this choice splits the viable set into two subsets. Because positions are correlated, the identity of this first residue constrains which amino acids are admissible at other positions and, crucially, it constrains them differently in the two subsets. Within each subset one can repeat the argument at a further position, splitting it again, where both the relevant position and its admissible choices may themselves differ between the two branches. Iterating, correlation among positions carves the viable set into nested subsets: a hierarchy. This is precisely the pattern a branching history would also imprint, which is the problem this paper addresses.

\begin{figure*}[t]
        \centering
        \includegraphics[width=7in]{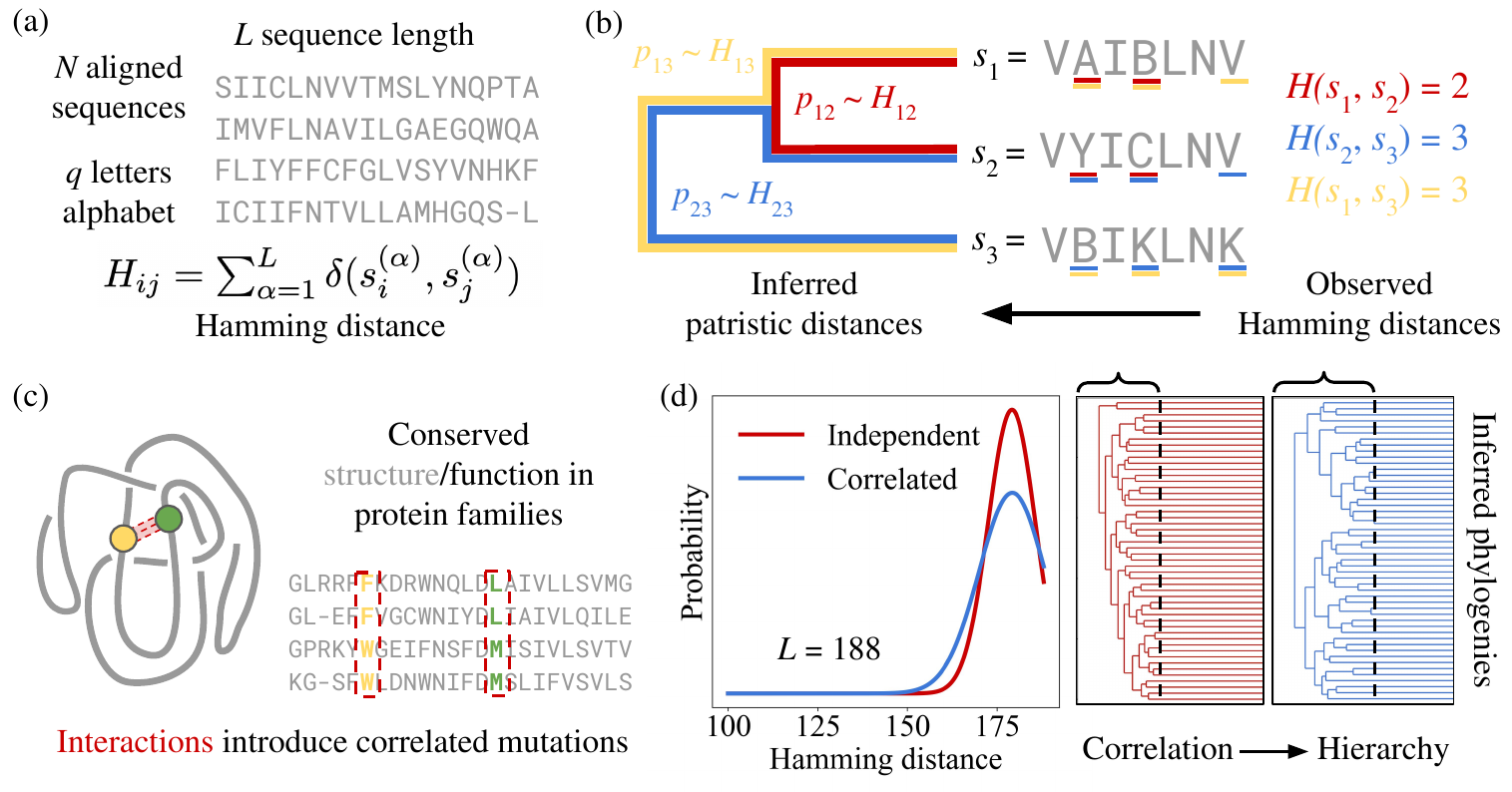}
        \caption{{\bf Distance-based phylogenetic inference from protein MSAs.} 
(a) Multiple sequence alignment (MSA) of observed protein sequences. The alignment maps variable-length proteins belonging to the same functional family into fixed-length discrete vectors. The number of mismatching positions between any pair of sequences defines the Hamming distance in this space. 
(b) Distance-based algorithms assume that the Hamming distance matrix reflects solely the branching history of the family, aiming to reconstruct a phylogenetic tree whose branch lengths (patristic distances) best reproduce the observed Hamming distances. 
(c) The three-dimensional fold of a protein introduces physical interactions that constrain the space of viable sequence variants: non-compatible residues in close spatial proximity destabilize the tertiary structure and impair biological function. These epistatic constraints manifest as multi-site correlations across the MSA. 
(d) Introducing coevolutionary correlations during sequence sampling distorts observed distance metrics and biases the hierarchy of reconstructed trees; stronger epistasis drives the inference toward artificially hierarchical branching patterns.}
        \label{fig:model}
\end{figure*}

When the most recent common ancestor of two genes lies relatively close in evolutionary time (the regime of shallow divergence) history and constraint can be treated as approximately separable. Sequences differ at comparatively few sites and one can plausibly attribute the pattern of differences to the branching order that produced them, reading constraint off as a slowly varying backdrop. Deep divergence dissolves this separation. In the ``twilight zone'', where pairwise identity falls toward the level expected between unrelated sequences~\cite{Doolittle1986,Rost1999}, the same sequence data used to infer phylogenetic relationships is simultaneously shaped by two ingredients that can no longer be cleanly disentangled: (i) the shared ancestry a tree is meant to encode, and (ii) the intrinsic hierarchy of the constrained subset of sequence space through which neutral evolution is free to roam~\cite{Rammal1986,Qin2018,RodriguezHorta2019,RodriguezHorta2021,Dietler2023,Colavin2022,Khatri2025,Weinstein2022}. Two distantly related proteins may look alike not because their lineages diverged recently but because the constraints admit only a limited repertoire of viable solutions, and both were driven into the same region of that repertoire.

Before we can read hierarchy in a set of sequences as evidence for a particular historical branching process, we must understand the hierarchy that is already present in the set of viable sequences favored by neutral evolution. Stated as a null hypothesis: even in the complete absence of a tree, {\it e.g.} for an ensemble of sequences sharing no genealogy beyond common membership in the same constrained set, nontrivial hierarchical structure can arise purely from the shape of the viable set. Any inference that maps such structure onto ancestry must first subtract what correlations contribute on their own.

In this paper, we explore this question using tools from statistical mechanics and multidimensional geometry to model the effect of structural constraints on the sequence space accessible to a protein family during neutral evolution. We show that, by mapping sequences into an appropriate metric space, epistasis can be naturally interpreted as a dimensionality reduction in sequence space. The resulting intrinsic dimension, imposed by a nontrivial free-energy landscape, encodes long-time dynamical properties of the family and allows us to estimate the probability that similarities observed in a multiple sequence alignment (MSA) arise by chance from the stochastic exploration of the family’s fitness landscape rather than from recent shared ancestry.

We first introduce and analyze the problem in a simple Ising-like toy model, which can be solved analytically. Within this framework, we show that increasing the strength of epistatic correlations in the sequences leads simultaneously to a reduction in the intrinsic dimensionality of sequence space, to changes in the hierarchical structure of the reconstructed phylogeny, and to an increase in the probability of observing small pairwise distances that arise from random encounters between independently evolving sequences. We refer to these events as recurrence events. Importantly, such events are not reliable proxies for recent shared ancestry, as they emerge from stochastic exploration of the constrained sequence landscape rather than from genealogical proximity.

We then apply our framework to a real protein family. To this end, we develop a method to systematically tune the degree of epistasis within the family and use Potts models together with Markov chain Monte Carlo simulations to generate sequences under controlled evolutionary constraints. We show that the intrinsic dimension inferred from the geometry of sequence space remains predictive of the recurrence probability of independently evolving sequences, even in the complex, multi-basins free-energy landscapes characteristic of real proteins.

\section{model setup}

We begin by defining the main tools and observables relevant to our phylogenetic inference problem. Consider a set of $N$ protein sequences belonging to a given protein family. These sequences are first processed through a Multiple Sequence Alignment (MSA), resulting in aligned sequences of identical length $L$~\cite{Durbin1998,Eddy2011,Katoh2013}. Each sequence is represented as a string of symbols drawn from an alphabet consisting of the 20 standard amino acids together with a gap character introduced during the alignment procedure. The set of sequences can be naturally viewed as a metric space by defining the Hamming distance
\begin{equation}
    H(s_i , s_j) = \sum_{\alpha=1}^{L} \delta\!\left(s_i^{(\alpha)}, s_j^{(\alpha)}\right),
\end{equation}

\noindent which counts the number of mismatched symbols between two sequences $s_i$ and $s_j$, and takes values in the range $[0, L]$.
In the simplest approaches, the Hamming distance is used as the input for inferring evolutionary relationships within a protein family: counting the number of mismatched positions between two sequences provides information about the number of evolutionary steps separating them in time. An analogous pairwise divergence signal underlies likelihood- and Bayesian-based reconstruction, which fit an explicit model of the substitution process to it rather than to the raw distances. Reconstructing the branching process that most faithfully reproduces the observed Hamming distances as branch lengths, also known as patristic distances, allows the Hamming distance matrix to be interpreted as the direct outcome of evolutionary dynamics alone, as shown in Fig.~\ref{fig:model}a,b.

This procedure implicitly makes two assumptions: (i) observed mutations are independent, and (ii) the probability of returning to an already visited sequence is negligible during the stochastic process underlying evolution, such that the Hamming distance does not decrease along the evolutionary trajectory. These assumptions are not peculiar to distance-based reconstruction. Assumption (i) is the site-independence assumption built into the standard substitution models used by maximum-likelihood and Bayesian methods, so correlated positions violate it irrespective of the estimator; assumption (ii) concerns the evolutionary process itself and therefore constrains any inference method, however sophisticated.

These assumptions are generally not satisfied in real protein families, where the three-dimensional fold associated with the family introduces physical contacts that constrain the allowed pairs of amino acids able to sustain such interactions and hence mutations in those sites are strongly correlated. This effectively reduces the number of degrees of freedom, making the accessible sequence space smaller than the nominal exponential space of size $21^L$ (see sketch in Fig.~\ref{fig:model}c).

The presence of such correlations has a non-trivial effect on the Hamming distance distribution. To illustrate this point, Fig.~\ref{fig:model}d compares the distribution obtained from independent sequences with that generated by sequences drawn with artificially correlated mutations, in which a subset of $L/2 = 94$ positions are constrained to mutate concomitantly in order to preserve prespecified pairwise contacts. For example, an F at position $i$ is always associated with an L at position $j$, while a W at position $i$ always implies an M at position $j$, as sketched in the MSA in Fig.~\ref{fig:model}c.

The distribution corresponding to the correlated sequences appears broader than that of the independent sequences, and the reconstructed phylogeny shown on the right exhibits a deeper hierarchical branching structure, arguably because of the need to accommodate a more heterogeneous set of distances, which the reconstruction algorithm (UPGMA in this case~\cite{Sokal1958,Felsenstein2004}) can only resolve by introducing nested branching patterns; the broadening of the underlying distance distribution, rather than the particular algorithm, is what forces the nesting. Note that the $x$-axes of the inferred phylogenies are normalized to make the comparison meaningful.

These observations suggest that standard phylogenetic inference is sensitive to two effects introduced by epistasis. On one hand, the presence of correlations effectively reduces the dimensionality of the accessible sequence space and mitigates the concentration-of-measure phenomenon, making the Hamming distance distribution less concentrated around its mean. On the other hand, the reduced dimensionality exponentially suppresses the number of accessible states, potentially making the probability of returning to an already visited sequence non-negligible.

In the following sections, we show that these intuitions can be formalized using a simple Ising-like model, within which both the intrinsic dimensionality and the recurrence probability can be quantified. The results obtained in this simplified setting are then extended to a realistic case study, providing a predictive framework for real protein families.

\begin{figure}[t]
        \centering
        \includegraphics[width=3.5in]{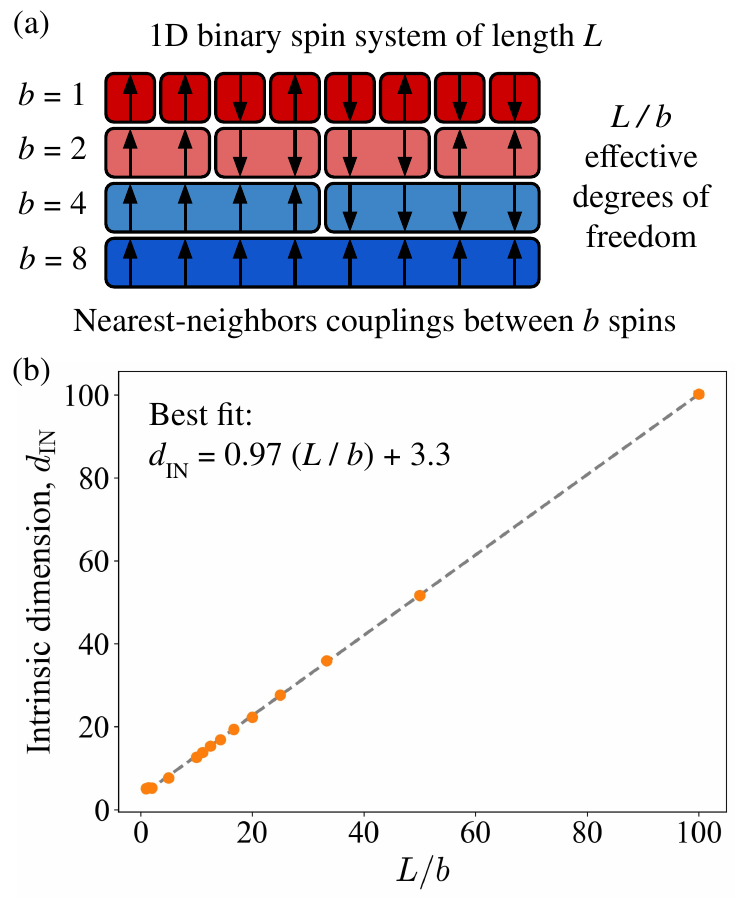}
        \caption{{\bf Controlled correlations in an Ising-like toy model.} 
(a) Using a two-letter alphabet, sequence space maps directly onto a one-dimensional spin system, where the interaction range of nearest-neighbor Hamiltonian couplings can be tuned to control equilibrium correlations among typical sequences. The block-wise structure of the interaction matrix renders the effective degrees of freedom inversely proportional to the block size $b$. The 1D Ising-like Hamiltonian admits an exact analytical solution, elucidating the precise link between microscopic interactions and geometric constraints. 
(b) Intrinsic dimension $d_{\rm IN}$ [Eq.~\eqref{eq:dIN}] as a function of the number of blocks $L/b$. Under the chosen parameters, this relationship is linear, with a slope close to 1 and a small vertical offset for large blocks arguably due to the short range nature of the interactions. Errors are comparable to the symbols size.}
        \label{fig:ising_model}
\end{figure}

\begin{figure*}[t]
        \centering
        \includegraphics[width=7in]{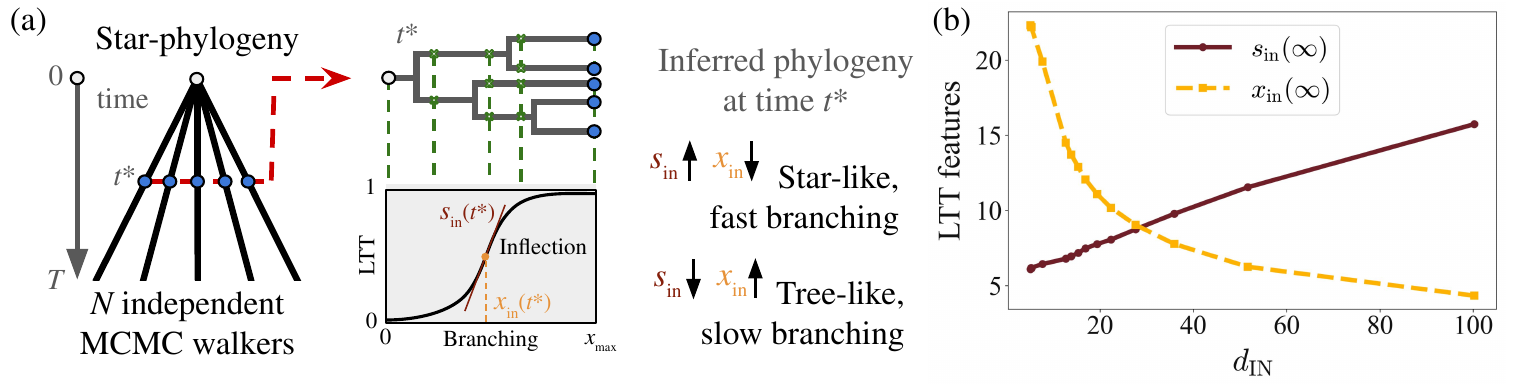}
       \caption{{\bf Effect of intrinsic dimension on the hierarchy of inferred phylogenies.} 
(a) Star-phylogeny simulations, initialized from a single common ancestor across all trajectories, allow us to test the inferred equilibrium hierarchy against a known ground truth lacking phylogenetic structure. Slicing trajectories at distinct time points and reconstructing the corresponding trees reveals how the inferred branching pattern evolves over time (see SI for details). The degree of hierarchy is quantified using lineage-through-time (LTT) curves, where steeper sigmoidal slopes ($s_{\rm in}$) and earlier inflection points ($x_{\rm in}$) indicate less hierarchical, more star-like tree topologies. 
(b) Equilibrium inflection coordinates $x_{\rm in}$ and slopes $s_{\rm in}$ plotted as a function of the intrinsic dimension $d_{\rm IN}$. Errors are comparable to the symbol size.}
        \label{fig:ising_results}
\end{figure*}

\section{A solvable Ising-like model for controlled correlation}

We start by formalizing the effect of correlation on the sequence space dimension using a simple toy model. To keep our model transparent and solvable, we focus on binary sequences composed of only two letters, A and B, corresponding to spin-up and spin-down states, respectively. Correlations between sequence positions are introduced via block structures of varying sizes, as illustrated in Fig.~\ref{fig:ising_model}a. Each block consists of sites coupled through nearest-neighbor ferromagnetic interactions, favoring uniform states within the block (either all A or all B). No interactions are present between blocks, ensuring that different regions of the sequence remain independent.
By varying the block size, we systematically control the degree of correlation within the sequence. When the block size is set to 1, spin flips (mutations) occur independently at each site, reproducing the traditional independent-site model. As the block size increases, intra-block correlations dominate the system’s evolution and the number of blocks $L/b$ can be interpreted as an effective number of independent degrees of freedom. This model thus provides a controlled framework for studying how varying levels of correlation influence the distribution of sequence distances and the resulting phylogenetic tree structure.

We simulated the evolution of spin sequences composed of 100 spins using a Markov Chain Monte Carlo (MCMC) algorithm with the Glauber acceptance rule, which ensures stationary sampling from the equilibrium Boltzmann distribution while maintaining local mutation dynamics and thermal coupling to the bath. We used a reduced temperature of $0.5$, which allowed us to cleanly measure the observables  of interest for all the explored block sizes. The block size was varied systematically over the range [1:100], see the Supporting Material for further details.

From a geometrical perspective, spin sequences can be represented as the vertices of an $L$-dimensional hypercube, and the variance of the normalized Hamming distance across all pairs of vertices scales as $\sigma^2_H = 1/(4L)$. This is a simple example of the well-known \textit{concentration of measure} phenomenon, in which, as the dimensionality of the space increases, the distribution of distances between points within a bounded region becomes sharply peaked around its mean value.

This observation suggests that the spreading of the distance distribution can serve as a direct measure of the intrinsic dimension of a dataset residing in a bounded metric space, such as the one defined by sequences and the Hamming distance~\cite{Facco2019}.

Motivated by the observed broadening of the distribution with increasing correlation, which effectively reduces the number of available degrees of freedom, and recalling that for independent spins the variance is inversely proportional to the system size, we define an \textit{intrinsic dimension} for our system as:
\begin{equation}
    d_{\rm IN} = \frac{1}{4\sigma^2_H}.
    \label{eq:dIN}
\end{equation}

Fig.~\ref{fig:ising_model}b demonstrates that the intrinsic dimension defined in Eq.~\ref{eq:dIN} scales linearly with the nominal degrees of freedom of the spin model. Specifically, we find a proportionality constant of 0.97 alongside a minor residual vertical shift of approximately 3.3. Furthermore, the expression for $d_{\rm IN}$ is geometrically consistent with the variance of the geodesic distance between points uniformly distributed on a hypersphere (see Supporting Information for mathematical details). Given that the sequence space of real proteins similarly constitutes a bounded metric space governed by the Hamming distance, we will exploit this geometric analogy to define an intrinsic dimension based on an abstract geometric embedding of the neighborhood graph derived from the observed MSA~\cite{Granata2016}.

To validate the robustness of this definition, we examine its relationship with the hierarchical structure of the phylogenetic tree inferred from the pairwise distance matrix of the sequences used for the $d_{\rm IN}$ calculation. To this end, we considered a ``star-phylogeny'' simulation consisting of 100 independent walkers, all originating from the same initial sequence, and tracked their evolution along their Monte Carlo (MC) trajectories. For each timestep $t^*$, we extracted the set of sequences from the independent walkers and inferred the corresponding tree using UPGMA (see sketch in Fig.~\ref{fig:ising_results}a). 

We then quantified the degree of hierarchy of the resulting tree using the Lineages Through Time (LTT) plot~\cite{Nee1992,Harvey1994,Nee1994}, which measures the fraction of active lineages as a function of branch depth in the inferred phylogeny. Recall that the branch depth in the LTT plot is entirely decoupled from the variable $t^*$, which instead measures the elapsed MC simulation time of the walkers from their common ancestor.

The LTT curve exhibits a characteristic sigmoidal shape, bounded between 0 and 1 on  the $y$-axis, and between 0 and the maximum pairwise distance of the matrix on the $x$-axis. From this curve, we extracted the $x$-coordinate of the inflection point, $x_{\rm in}$, and the slope of the tangent line at this same point, $s_{\rm in}$. These two independent metrics quantify how early the branching events occur within the tree and how rapidly the lineage accumulation saturates, respectively. Consequently, a tree characterized by a lower $x_{\rm in}$ and a higher $s_{\rm in}$ is less hierarchical (i.e., more star-like), and vice versa.

Fig.~\ref{fig:ising_results}b shows the equilibrium values of $s_{\rm in}$ and $x_{\rm in}$, obtained after the MC simulation has reached its steady-state distribution, as a function of the intrinsic dimension ($d_{\rm IN}$) of the sequence set. The corresponding time trajectories are provided in the Supporting Information. The slope at the inflection point increases monotonically with $d_{\rm IN}$, exhibiting a threefold increase as $d_{\rm IN}$ varies from 5 to 15; conversely, the horizontal position ($x$) of the inflection point decreases from above 20 to below 5. Note that, because the LTT curves quantify the fraction of active lineages along the tree reconstructed from the Hamming distance input matrix, both $s_{\rm in}$ and $x_{\rm in}$ are dimensionless quantities.
These results confirm that as the intrinsic dimension of the sequence set increases, the tree inferred from the Hamming distance measured in the equilibrium ensemble becomes less hierarchical and more closely reflects the star phylogeny ground truth. It is important to emphasize that all the phylogenetic relationships inferred in this experiment are merely a consequence of the correlation structure of the state space and are completely unrelated to actual shared ancestry, which is entirely absent in our setting. Our observations underscore the fact that Hamming distances contain information that stems not only from evolution, but also from the physical constraints imposed on the accessible sequences.

Finally, we evaluated how $d_{\rm IN}$ predicts the long-term dynamics of our spin model. For this purpose, we invoke Kac's lemma from dynamical systems theory, which states that for a subset $A$ of the phase space with equilibrium measure $\pi(A)$, the average recurrence time $\tau_{\rm rec}(A)$ for a trajectory initiated in $A$ to return to the set is inversely proportional to its equilibrium probability, $\tau_{\rm rec}(A) \propto 1/\pi(A)$~\cite{Kac1947}. 

The equilibrium measure of a set is given by the ratio between the restricted partition function and the total partition function of the system, $\pi(A) = Z_A / Z$. Considering $A$ as a ball of small radius $R$ centered around a typical initial sequence, we can approximate the restricted partition function by the volume of the ball. Using the exact expression of $Z$ for an Ising-like system, we predict that 
\begin{equation}
    \log \tau_{\rm rec} \sim k(R) \cdot d_{\rm IN} + \text{const.},
    \label{eq:rec_ising}
\end{equation}
where $k(R)$ is a radius-dependent slope and the constant is independent of $R$ (see SI for analytical and numerical details).

We numerically validated this prediction by measuring the recurrence time of a trajectory to a ball of radius $R$, centered on a reference sequence, following its exit. We generated $100$ independent Markov chain trajectories initiated from canonical sequences and computed their mean recurrence time.

\begin{figure}[t]
        \centering
        \includegraphics[width=3.5in]{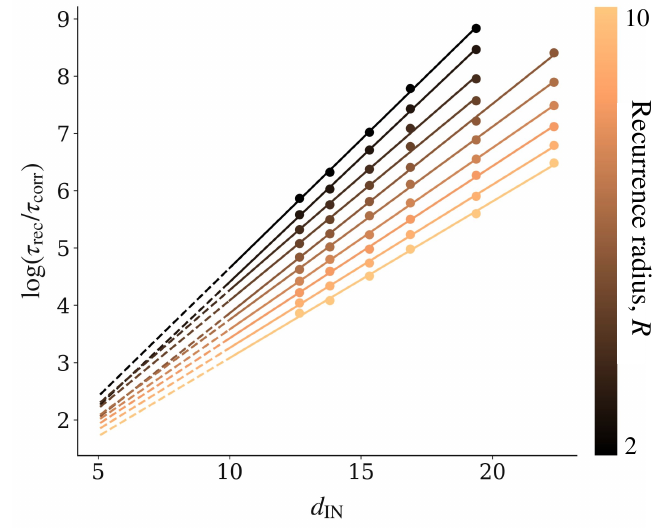}
      \caption{{\bf Recurrence time as a function of the intrinsic dimension.} 
Logarithm of the rescaled recurrence time $\log(\tau_{\rm rec} / \tau_{\rm corr})$ into a ball of radius $R$, plotted as a function of the intrinsic dimension $d_{\rm IN}$ of the dataset. The linear scaling predicted by Eq.~\eqref{eq:rec_ising} is well reproduced by the data across all tested radii $R$. Although fitted independently, the linear extrapolations converge toward a common intercept, in agreement with theoretical expectations. Decreasing the radius $R$ exponentially suppresses the recurrence probability, rendering recurrence events computationally inaccessible at small radii within accessible simulation timescales. Error bars are comparable to the symbol size.}
        \label{fig:ising_recurrence}
\end{figure}

Figure~\ref{fig:ising_recurrence} presents $\log \tau_{\rm rec}$ as a function of $d_{\rm IN}$ for several recurrence radii $R$. Here, the integrated autocorrelation time $\tau_{\rm corr}$ of the Markov chain was used to rescale the recurrence times, controlling for the intrinsic dynamical timescales introduced by different Hamiltonian parameters and for the specific acceptance condition of the chain.

These results demonstrate that the predicted linear scaling holds robustly across different values of $d_{\rm IN}$ over the computationally accessible range of recurrence radii ($R \in [\text{2}, \text{10}]$); although the intercepts were fitted independently for each radius $R$, the extrapolated lines converge to a common value, in agreement with Eq.~\eqref{eq:rec_ising}. Crucially, decreasing the intrinsic dimension by less than a factor of two reduces the recurrence time by two to four orders of magnitude, depending on the chosen radius. This confirms that $d_{\rm IN}$ captures long-term dynamics, providing a quantitative metric to evaluate whether small sequence distances reflect evolutionary reconvergence within an epistatically constrained landscape rather than recent shared ancestry. Note that, to a first approximation, the recurrence time sets the timescale over which the probability of recurrence exponentially saturates to unity. This effectively identifies a threshold in root-to-tip tree depth beyond which small Hamming distances can no longer serve as a reliable proxy for shared ancestry.

\section{natural protein family}

\begin{figure*}[t]
        \centering
        \includegraphics[width=7in]{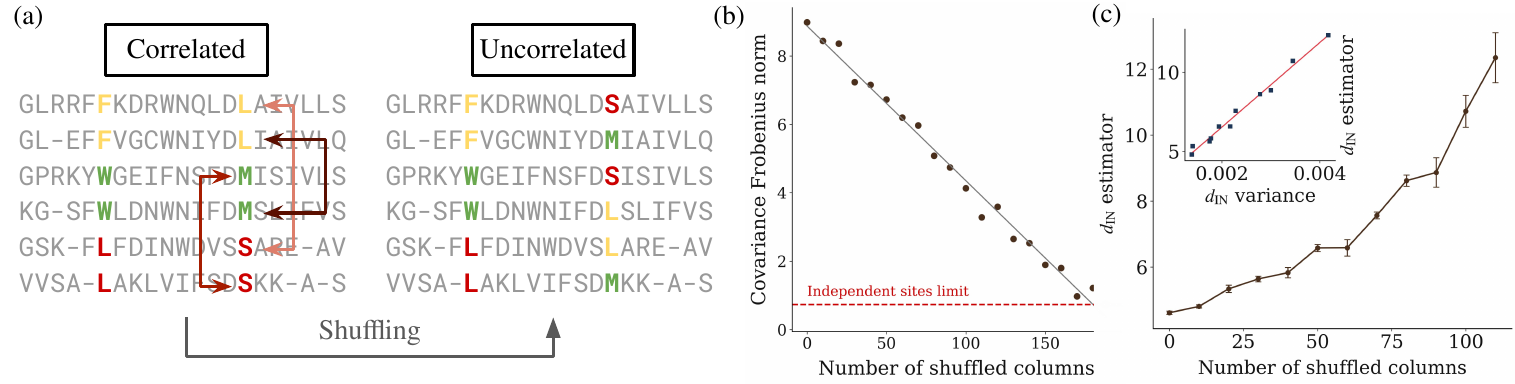}
\caption{{\bf Shuffling procedure and intrinsic dimension in natural proteins.} 
(a) To systematically tune the epistatic couplings in a natural protein MSA, we selected random subsets of columns and permuted the amino acid residues within each column, thereby destroying pairwise correlations while preserving single-site frequencies. 
(b) The norm of the covariance matrix across varying degrees of shuffling confirms that the procedure systematically degrades correlations, smoothly approaching the independent-site limit (see SI for details). 
(c) The intrinsic dimension $d_{\rm IN}$ plotted as a function of the number of shuffled columns exhibits a monotonic increase, demonstrating that suppressing epistasis expands the effective geometric dimension of the sequence space. The inset confirms that the direct evaluation via Eq.~\eqref{eq:dIN} is proportional to the geodesic-based estimator developed in Ref.~\cite{Granata2016}, differing only by an overall scaling factor.}

        \label{fig:potts_ID}
\end{figure*}

Building on the synthetic results from the previous section, we now analyze a real biomolecular system to demonstrate that our framework extends naturally to complex, biologically relevant settings. We consider the PFAM family PF00520 ($L = 188$, $q = 21$, corresponding to the 20 standard amino acid types plus a gap character). Sequences in this family are mapped to the vertices of an hypercube in a high-dimensional discrete representation via one-hot encoding. Specifically, each amino acid symbol is represented as a canonical basis vector in $\mathbb{R}^q$ (e.g., $\text{A} \to (1,0,\dots,0)$, $\text{C} \to (0,1,\dots,0)$), such that a sequence of length $L$ maps to a sparse vector of dimension $Lq$, subject to the constraint that exactly one component equals unity in each $q$-dimensional block~\cite{Morcos2011,Cocco2018}. 
The natural multiple sequence alignment of the family is first filtered to remove sequences with a normalized Hamming distance $p < 0.2$, and columns with a gap fraction exceeding $80\%$, thereby eliminating signals from recent shared ancestry and uninformative positions. We then infer an effective Potts Hamiltonian matching the empirical one- and two-point sequence statistics via a MCMC inverse inference approach~\cite{Haldane2021}.
The resulting energy landscape features a complex network of interactions between amino acid positions, yielding a multi-basins, spin-glass-like free-energy in the thermodynamic limit, and captures the coevolutionary correlations observed in natural sequences via pairwise couplings. This provides an effective energetic surface that constrains evolutionary trajectories~\cite{Weigt2009,Levy2017,Hopf2017,delaPaz2020,Bisardi2022,DiBari2024}.

To systematically tune the two-point correlations in the resulting Hamiltonian, and hence the epistatic constraints of the family, we implemented a column-wise shuffling procedure on the MSA, as illustrated in Fig.~\ref{fig:potts_ID}a. The algorithm randomly selects a subset of columns in the MSA and randomly permutes the residues within each chosen column. This destroys the pairwise correlations between the selected columns and the rest of the alignment while preserving single-site amino acid frequencies. Following the shuffling step, we re-applied the MCMC inference to fit a new coupling matrix with a reduced degree of correlation.
Figure~\ref{fig:potts_ID}b shows the effect of the shuffling procedure on the norm of the covariance matrix of the MSA, confirming that it linearly degrades the correlation structure and recovers the independent-site limit when all columns are randomized (see SI for details).

Having established a parameter to systematically tune the correlation structure, we measured the intrinsic dimension $d_{\rm IN}$ of the datasets as a function of the number of shuffled columns. As expected, this yielded an increase in dataset dimension as the degree of shuffling increased (i.e., as correlations decreased).

However, we encountered two key differences compared to the previous model: (i) direct application of the formula in Eq.~\ref{eq:dIN} resulted in fractional intrinsic dimension values, preventing a clear geometric interpretation of the number without an appropriate rescaling factor; and (ii) interpreting the set of sequences as vertices of an hypercube conflates constraints imposed by true physical correlations with spurious ones introduced by the one-hot encoding scheme (i.e., $\sum_{k=1}^q \sigma_k(i) = 1, \; \forall i \in [1, L]$).

To confirm the robustness of our definition in this broader context, we also employed the intrinsic dimension estimator developed in Ref.~\cite{Granata2016}. This method constructs the nearest-neighbor graph for a given set of points (the one-hot encoded sequences) using a specified metric (the Hamming distance) and fits the graph distances along spherical geodesics to infer the optimal dimension of the embedding sphere. While a spherical embedding restricts curvature variability across sequences, it provides a global, family-wide estimate of the spatial dimension required to accommodate the observed sequences. To further confirm the robustness of our definition, we computed $d_{\rm IN}$ from Eq.~\ref{eq:dIN} using a set of uncorrelated sequences obtained from long simulations of the fitted Potts model to remove any residual signal of phylogeny that may be present in the natural MSA; we also used a polytope-based estimator designed for discrete spaces~\cite{Macocco2023}, obtaining coherent results between methods (see SI for details).

The results are presented in Fig.~\ref{fig:potts_ID}c, where the geodesic $d_{\rm IN}$ is plotted against the number of shuffled columns. Notably, the upward trend is preserved, the numerical scale on the $y$-axis now admits a direct interpretation as a spatial dimension, and the two estimators display a linear dependence (see inset), confirming that both capture the same underlying geometric feature of the dataset.
The magnitude of $d_{\rm IN}$ obtained here is consistent with a growing body of independent evidence that protein sequence space is far lower dimensional than its nominal size. A small intrinsic dimension was first reported for protein families using graph-distance estimators~\cite{Granata2016}, subsequently confirmed with neighborhood-based estimators~\cite{Facco2019} and with an estimator formulated directly for discrete metrics~\cite{Macocco2023}, and most recently recovered proteome-wide using the correlation dimension~\cite{Grassberger1983,Isakova2026}. That estimators resting on different geometric assumptions agree on values orders of magnitude below the nominal dimension $Lq$ indicates that the low dimensionality is a property of the data rather than an artifact of any particular construction.

\begin{figure}[t]
        \centering
        \includegraphics[width=3in]{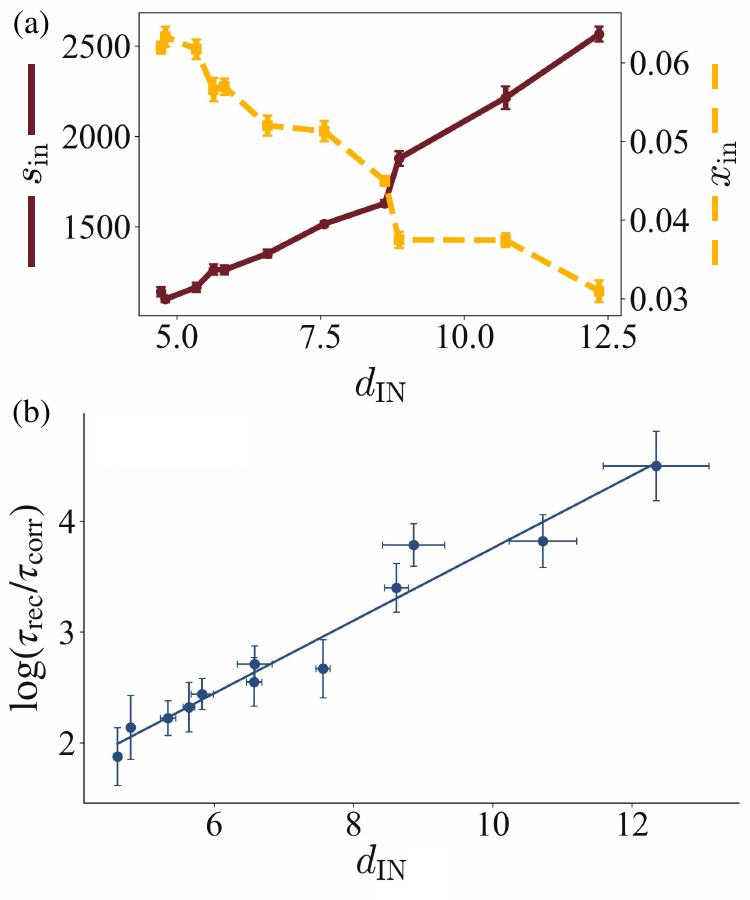}
\caption{{\bf Branching and recurrence in natural protein families.} 
(a) Lineage-through-time (LTT) parameters as a function of the intrinsic dimension $d_{\rm IN}$. Larger slopes ($s_{\rm in}$) and smaller inflection time coordinates ($x_{\rm in}$) are observed as $d_{\rm IN}$ increases, indicating a more star-like reconstructed tree. 
(b) Logarithm of the rescaled recurrence time $\log(\tau_{\rm rec} / \tau_{\rm corr})$ as a function of $d_{\rm IN}$ for a recurrence radius $R = 112$ ($0.6L$). The scaling shows an approximately linear increase despite the complex, spin-glass-like underlying dynamics of sequence evolution. Due to the high nominal dimensionality of the sequence space ($L = 188, q = 21$), smaller recurrence radii exponentially suppress the recurrence probability, making it impossible to collect recurrence statistics within computationally accessible simulation timescales, see SI for more details.}
        \label{fig:potts_recurrence}
\end{figure}

We then evaluated whether the intrinsic dimension remains predictive of branching patterns reconstructed by distance-based inference algorithms (UPGMA) in this more general setting. To this end, we repeated the star-phylogeny experiment designed for our toy spin model, initializing 100 independent trajectories from the same initial sequence and evolving them under the fitted Potts Hamiltonian. We then measured the slope and position of the lineage-through-time (LTT) inflection point in the long-time limit across different values of $d_{\rm IN}$, as shown in Fig.~\ref{fig:potts_recurrence}a. 

Consistent with our previous results, increasing the intrinsic dimension yields a larger slope ($s_{\rm in}$) and a smaller time coordinate ($x_{\rm in}$) for the LTT inflection point. This demonstrates that the hierarchy of the reconstructed tree shifts according to the degree of equilibrium correlation present in the sequences, thereby distorting the phylogenetic signal one aims to reconstruct. Across the observed $d_{\rm IN}$ range, the branching velocity in the inferred tree ($s_{\rm in}$) increases by a factor of $2.5$, while the onset of branching ($x_{\rm in}$) decreases by a factor of $2$. This indicates that a larger $d_{\rm IN}$ produces a more star-like reconstructed tree, with branching occurring earlier and at an accelerated pace.

We finally performed extended MCMC simulations to track recurrence events during evolution across the different landscapes; the results are presented in Fig.~\ref{fig:potts_recurrence}b. Although the dynamics of the system appear considerably more complex than in the toy model, as evidenced by the metastable behavior of the Hamming distance, which can be attributed to the spin-glass-like character of the Potts Hamiltonian (see SI for details), the logarithm of the recurrence time depends linearly on the estimated intrinsic dimension, increasing by a factor of $2$ across the observed range. The applicability of Kac's lemma in this more general context is also confirmed by the linear relation between the logarithm of $\tau_{\rm rec}$ and the logarithm of the probability mass concentrated in Hamming distances smaller than the recurrence radius $R$, suggesting that the intrinsic dimension is indeed informative of the volumes of recurrence balls and that this volume controls the recurrence probability of the sequence; see SI for details. This confirms that even when the nominal dimension of the sequence space is large, strong epistasis can exponentially compress the effective spatial dimension, thereby confounding distance-based phylogenetic algorithms when reconstructing the true evolutionary signal.

\section{Discussion and Conclusions}

In this work, we have investigated how epistatic constraints influence the hierarchical structure of phylogenies reconstructed via distance-based methods. We demonstrated that coevolutionary correlations between amino acid positions alter the pairwise Hamming distance distribution, systematically biasing the tree topologies inferred by distance-matrix algorithms such as UPGMA.

To establish a transparent geometric picture of this phenomenon, we first introduced a one-dimensional Ising-like toy model in which the number of accessible degrees of freedom can be tuned analytically as a function of the correlation strength. Within this framework, we defined an intrinsic dimension $d_{\rm IN}$ based on the variance of the Hamming distance distribution, exploiting the concentration of measure phenomenon that governs distance distributions as spatial dimensionality increases. Under this formulation, stronger epistatic couplings directly manifest as a compression of the effective spatial dimension. Using this metric, we established two key results: (i) $d_{\rm IN}$ governs the hierarchical structure of reconstructed trees (quantified via lineage-through-time curves), where higher intrinsic dimensionality (weaker epistasis) leads to accelerated, star-like branching topologies and, (ii) $d_{\rm IN}$ predicts the long-time recurrence dynamics of sequence evolution, where reduced dimensionality exponentially increases the probability of sequence reconvergence at small Hamming distances, thereby undermining the foundational assumption that short sequence distances reflect recent shared ancestry.

We subsequently extended our framework to natural protein families modeled by inverse Potts Hamiltonians. Despite the complex, multi-basin energy landscapes and metastable dynamics characteristic of real proteins, our findings remain remarkably robust: $d_{\rm IN}$ consistently captures the spatial compression induced by epistasis and remains predictive of both phylogenetic tree distortions and long-time recurrence rates.

Our results bear on a recent proteome-wide analysis that reaches the same geometric conclusion but a different causal one. Isakova \emph{et al.}~\cite{Isakova2026} report effective dimensions typically below ten across thousands of vertebrate and bacterial families, orders of magnitude below the theoretical maximum, and attribute this compression primarily to descent from a single common ancestor, a ``point of origin'' effect, concluding that epistasis is globally permissive and cannot substantially reduce the dimensionality of extant sequences. The shuffling experiment reported here is a direct test of that conclusion and does not support it. Because independent equilibrium samples are drawn from a model distribution before and after shuffling, no phylogeny of any kind enters either ensemble, and the resulting change in $d_{\rm IN}$ is attributable to the two-point structure of the distribution alone; we find that this structure compresses the dimension substantially. The discrepancy is traceable to how epistasis is represented. In Ref.~\cite{Isakova2026} the viable set remains a Cartesian product of site-wise alphabets at every instant, with epistasis implemented as a stochastic reallocation of which states are allowed, so that the product structure is preserved and only its composition drifts. Such a model cannot generate a persistent low-dimensional manifold, and therefore cannot test whether epistasis of the kind imposed by folding, packing and binding does so; a Potts landscape, whose viable set is irreducibly non-product, can. We stress that this argument concerns the mechanism and not the attribution. Our couplings are inferred from a natural alignment and consequently absorb whatever correlations that alignment contains, phylogenetic ones included~\cite{Qin2018,RodriguezHorta2019,RodriguezHorta2021,Dietler2023,Khatri2025}, so the decomposition of the observed $d_{\rm IN}$ of a natural family into constraint and ancestry remains open.

A second line of evidence counsels caution about when the recurrence regime is actually entered. Ancient families appear not to have equilibrated: protein families already present in the last universal common ancestor are still diverging, a slow approach attributed to the fact that most sites become substitutable only after compensatory changes elsewhere~\cite{Povolotskaya2010}. Consistent with this, Ref.~\cite{Isakova2026} estimates that the dimensionality of simulated families saturates only once the tree accumulates on the order of nine substitutions per site, against roughly one substitution per site since the last universal common ancestor. Our analysis does not settle whether real families have crossed the recurrence horizon, and the two quantities that set it move in opposite directions: strong epistasis lowers $d_{\rm IN}$, which shortens $\tau_{\rm rec}$, but it also slows the dynamics, which lengthens $\tau_{\rm corr}$. Because recurrence times are reported here in units of $\tau_{\rm corr}$, this trade-off is not visible in Fig.~\ref{fig:potts_recurrence}b, and the absolute horizon may well lie beyond the age of the deepest extant families. We regard this as an open question. Expressing $\tau_{\rm rec}$ in substitutions per site and comparing it with the divergences at which phylogenetic inference is actually attempted is the natural way to settle it.

Whether molecular convergence exceeds neutral expectation remains contested, with reported excesses~\cite{Rokas2008,Goldstein2015} failing to survive nulls that assign site-specific sets of acceptable amino acids~\cite{Zou2015,Thomas2015,Storz2016}; that literature, however, measures a different quantity. Convergent and parallel substitutions are counted per site between reconstructed ancestral states, and only at positions where both lineages substituted, so conserved positions are excluded by construction and correlations between positions enter the count only indirectly. Recurrence, by contrast, is a property of whole sequences: it is the chance that two independently evolving lineages fall within a small Hamming distance of one another, rather than a coincidence at an individual position. By Kac's lemma it is set by the equilibrium mass $\pi(A)$ of a ball in sequence space, that is, by the left tail of the pairwise distance distribution. Inter-site correlations are what control this mass, the mean Hamming distance being fixed by the single-site frequencies alone, and $d_{\rm IN}$ [Eq.~\eqref{eq:dIN}] is what predicts it. The two sets of observations are thus complementary.

Our results highlight the necessity of evaluating reconstructed phylogenies through the lens of underlying epistatic constraints, demonstrating that patristic distances between observed sequences must be interpreted relative to the physical interaction network of the protein family. Looking forward, extending this framework will require accounting for spin-glass-like dynamics and landscape metastability to refine the geometric description of correlated sequence spaces. Crucially, mapping the recurrence timescales of sequence dynamics opens the possibility of establishing quantitative confidence thresholds for phylogenetic inference, enabling researchers to identify reliable branch depths as a function of family-specific epistatic strength and tree depth.

\bibliography{bibliography}

\clearpage
\onecolumngrid
\appendix 
\sffamily

\centerline{\Large Supporting Information}

\vskip 0.2cm
\centerline{\Large \shortstack{Recurrence in Sequence Space: How Epistasis Mimics Shared Ancestry}}
\vskip 0.2cm
\centerline{\large \shortstack{by P.~Chiarantoni, M.~Mahboubi, J.~Lamanna, V.~Carnevale}}
\vskip 1.5cm

\section{Ising-like model}
\subsection{Simulation details}
We performed Monte Carlo simulations of our toy model spin system using the Glauber acceptance rule. We set the total length of the chains to $L=100$, the temperature to $T=0.5$, and varied the block size $b$ within the set $b= [1;10],20,50,70,100$, passing from independent spins to a full 1D Ising model. 

We initialized 10 star phylogenies by first sampling 10 independent initial sequences uniformly. Then, from each sequence, we started 100 independent simulations, for a total of 1000 walkers for each value of $b$. We simulated each walker for a total of $10^6$ MC timesteps.

We used these star phylogenies to compute the LTT feature reported in the main text.

To obtain the equilibrium samples for each value of the block size $b$, we downsampled the second half ($5 \cdot 10^5$ steps) of the 1000 walkers using a stride of 100000 steps, based on the chain autocorrelation time (see following sections). This procedure resulted in a total of 5000 independent samples, which we then used to estimate the intrinsic dimension of the sequence space.

The 1-million-timestep trajectories were also used to quantify the recurrence probability and the related recurrence time by analyzing the Hamming distance trajectories from the initial sequence for each walker.

\subsection{Hamming distance}
\begin{figure*}[h]
        \centering
        \includegraphics[width=7in]{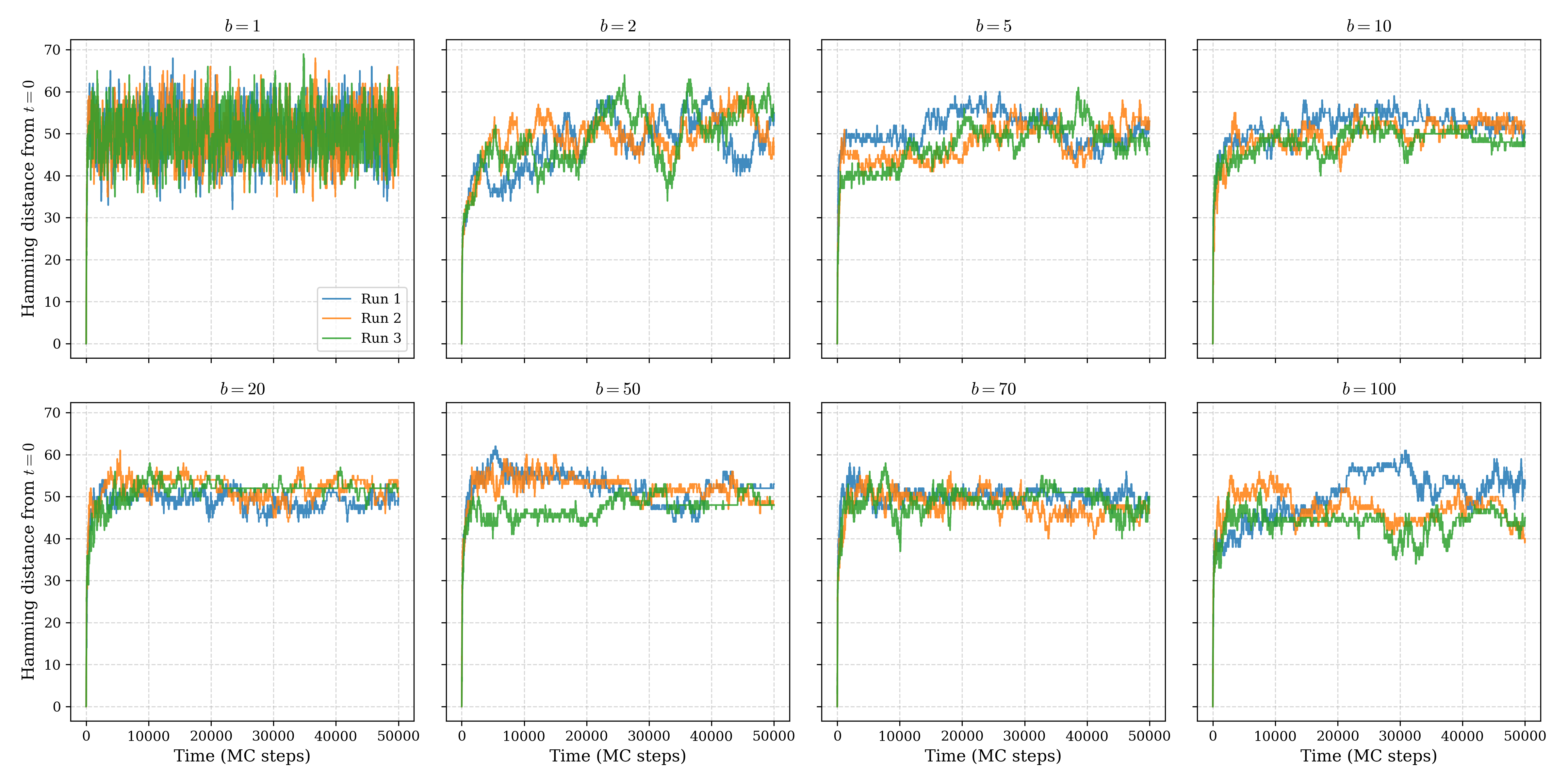}
       \caption{Trajectories of Hamming distances from the initial sequence for different values of the block size $b$.}
        \label{fig:SI_ising_hamming_traj}
\end{figure*}

\begin{figure*}[h]
        \centering
        \includegraphics[width=4in]{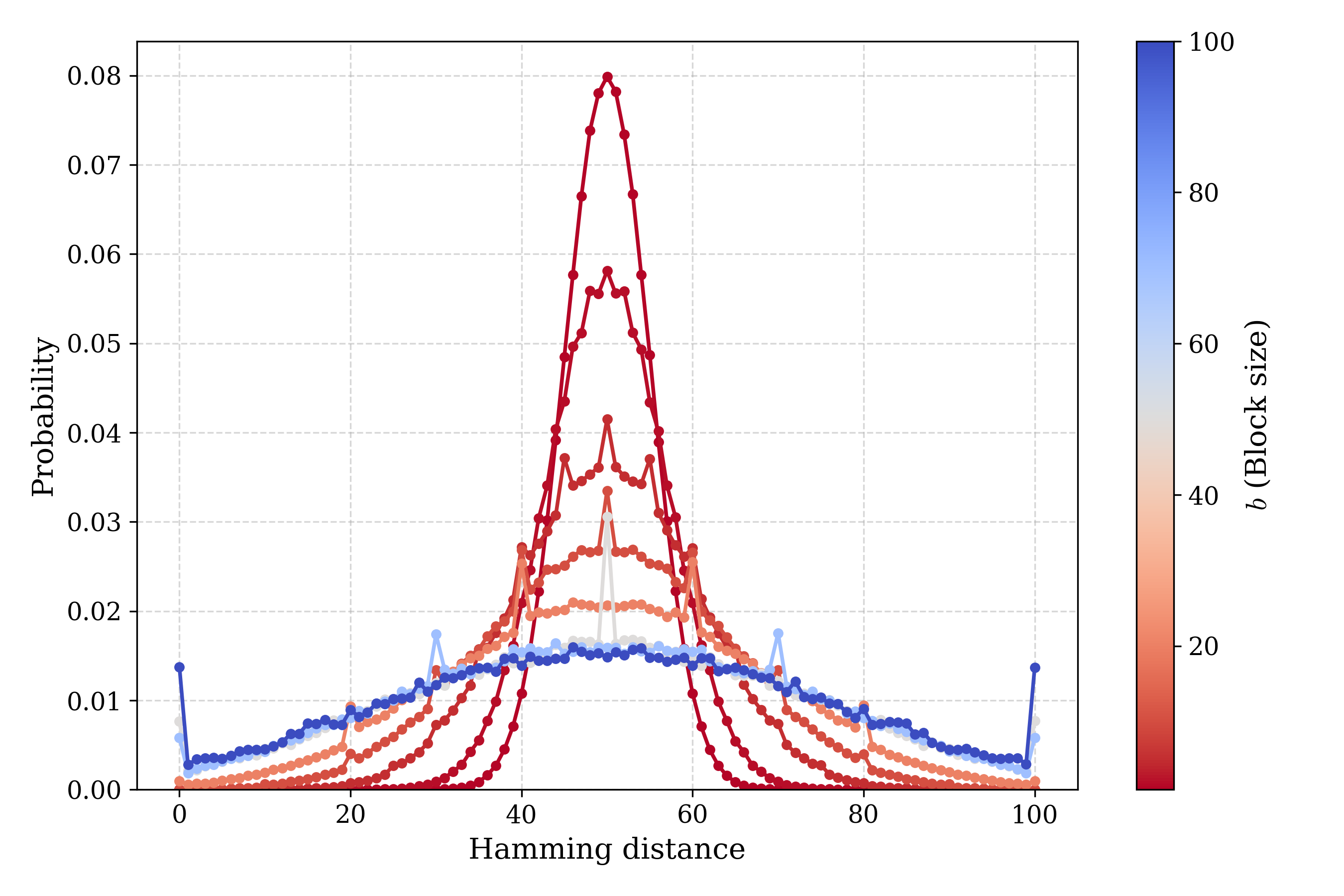}
       \caption{Probability distribution of the Hamming distances between pairs of canonical sequences for different values of the block size $b$.}
        \label{fig:SI_ising_hamming_ddp}
\end{figure*}

\subsection{$d_{\rm IN}$ analytical estimation}

We consider two independent spin configurations,
\(\mathbf{s}=(s_1,\ldots,s_L)\) and
\(\mathbf{t}=(t_1,\ldots,t_L)\), sampled from the equilibrium
Boltzmann distribution of the block-structured Ising model.
The spins take values \(s_i,t_i\in\{-1,+1\}\), and the Hamiltonian is
\begin{equation}
    \mathcal{H}(\mathbf{s})
    =
    -J\sum_{m=0}^{M-1}
    \sum_{j=1}^{b-1}
    s_{mb+j}s_{mb+j+1},
    \qquad
    M=\frac{L}{b}.
\end{equation}
We assume open boundary conditions within each block, no external
magnetic field, and \(J>0\).

The Hamming distance between the two configurations is defined as
\begin{equation}
    H(\mathbf{s},\mathbf{t})
    =
    \sum_{i=1}^{L} h_i,
    \qquad
    h_i=\frac{1-s_it_i}{2}.
    \label{eq:hamming_definition}
\end{equation}
Here, \(h_i=1\) if the spins disagree at site \(i\), and \(h_i=0\)
otherwise. We also introduce the normalized Hamming distance
\begin{equation}
    h=\frac{H}{L},
    \label{eq:normalized_hamming_definition}
\end{equation}
which takes values between zero and one.

Our objective is to derive the variance of the Hamming distance,
denoted by
\begin{equation}
    {\rm Var} (H)
    =
    \langle H^2\rangle-\langle H\rangle^2,
    \label{eq:hamming_variance_definition}
\end{equation}
and the variance of its normalized counterpart,
\begin{equation}
    \sigma_H^2
    =
    \langle h^2\rangle-\langle h\rangle^2
    =
    \frac{{\rm Var} (H)}{L^2}.
    \label{eq:normalized_hamming_variance_definition}
\end{equation}

Since the two configurations are sampled independently from the
same zero-field equilibrium distribution, spin-flip symmetry implies
\begin{equation}
    \langle s_i\rangle
    =
    \langle t_i\rangle
    =0.
\end{equation}
Moreover, independence of the two replicas gives
\begin{equation}
    \langle s_i t_i\rangle
    =
    \langle s_i\rangle\langle t_i\rangle
    =0.
\end{equation}
Therefore,
\begin{equation}
    \langle h_i\rangle
    =
    \frac{1-\langle s_i t_i\rangle}{2}
    =\frac{1}{2},
\end{equation}
and the mean Hamming distance is
\begin{equation}
    \langle H\rangle=\frac{L}{2}.
    \label{eq:hamming_mean}
\end{equation}

The variance of the sum in Eq.~\eqref{eq:hamming_definition} is
\begin{equation}
    {\rm Var} (H)
    =
    \sum_{i=1}^{L}\sum_{j=1}^{L}
    \operatorname{Cov}(h_i,h_j).
    \label{eq:hamming_variance_covariance}
\end{equation}
For each pair of sites,
\begin{align}
    \langle h_i h_j\rangle
    &=
    \frac{1}{4}
    \left\langle
        (1-s_it_i)(1-s_jt_j)
    \right\rangle
    \nonumber\\
    &=
    \frac{1}{4}
    \left[
        1-\langle s_it_i\rangle
         -\langle s_jt_j\rangle
         +\langle s_i s_j t_i t_j\rangle
    \right].
\end{align}
Independence of the two replicas implies
\begin{equation}
    \langle s_i s_j t_i t_j\rangle
    =
    \langle s_i s_j\rangle
    \langle t_i t_j\rangle
    =
    \langle s_i s_j\rangle^2.
\end{equation}
Since \(\langle s_it_i\rangle=0\), we obtain
\begin{equation}
    \langle h_i h_j\rangle
    =
    \frac{1}{4}
    \left[1+\langle s_i s_j\rangle^2\right],
\end{equation}
and hence
\begin{equation}
    \operatorname{Cov}(h_i,h_j)
    =
    \frac{1}{4}\langle s_i s_j\rangle^2.
    \label{eq:hamming_covariance}
\end{equation}
This expression also holds for \(i=j\), since
\(\langle s_i^2\rangle=1\) and
\(\sigma_{h_i}^2=1/4\).

Combining Eqs.~\eqref{eq:hamming_variance_covariance} and
\eqref{eq:hamming_covariance}, we obtain
\begin{equation}
        {\rm Var} (H)
        =
        \frac{1}{4}
        \sum_{i=1}^{L}\sum_{j=1}^{L}
        \langle s_i s_j\rangle^2.
    \label{eq:hamming_variance_general}
\end{equation}

Within each block, the system is a one-dimensional Ising chain with
nearest-neighbor coupling \(J\). Define
\begin{equation}
    q=\tanh(\beta J).
    \label{eq:ising_correlation_parameter}
\end{equation}
For an open, zero-field Ising chain, the equilibrium spin correlation
function is
\begin{equation}
    \langle s_i s_j\rangle
    =
    \begin{cases}
        q^{|i-j|}, & \text{if \(i,j\) belong to the same block},\\
        0, & \text{if \(i,j\) belong to different blocks}.
    \end{cases}
    \label{eq:block_spin_correlation}
\end{equation}
The first line follows from the fact that each bond variable
\(s_k s_{k+1}\) has mean \(q\), while the second follows from
independence of different blocks and the vanishing magnetization.

Substituting Eq.~\eqref{eq:block_spin_correlation} into
Eq.~\eqref{eq:hamming_variance_general}, only pairs of sites within
the same block contribute. Since all blocks are identical,
\begin{equation}
    {\rm Var} (H)
    =
    \frac{M}{4}
    \sum_{i=1}^{b}\sum_{j=1}^{b}
    q^{2|i-j|}.
    \label{eq:hamming_variance_block_sum}
\end{equation}
Separating diagonal and off-diagonal terms gives
\begin{equation}
    {\rm Var} (H)
    =
    \frac{M}{4}
    \left[
        b+2\sum_{r=1}^{b-1}(b-r)q^{2r}
    \right].
    \label{eq:hamming_variance_finite_sum}
\end{equation}
Introducing
\begin{equation}
    a=q^2=\tanh^2(\beta J),
    \label{eq:a_definition}
\end{equation}
we can write
\begin{equation}
        {\rm Var} (H)
        =
        \frac{L}{4b}
        \left[
            b+2\sum_{r=1}^{b-1}(b-r)a^r
        \right].
    \label{eq:hamming_variance_exact_sum}
\end{equation}

Using the finite-sum identity
\begin{equation}
    \sum_{r=1}^{b-1}(b-r)a^r
    =
    \frac{a\left[(b-1)-ba+a^b\right]}{(1-a)^2},
    \qquad a\neq 1,
\end{equation}
the exact variance becomes
\begin{equation}
        {\rm Var} (H)
        =
        \frac{L}{4b}
        \left[
            b+
            \frac{
                2a\left[(b-1)-ba+a^b\right]
            }{(1-a)^2}
        \right].
    \label{eq:hamming_variance_closed_form}
\end{equation}

From Eq.~\eqref{eq:normalized_hamming_variance_definition}, the
variance of the normalized Hamming distance is
\begin{equation}
    \sigma_H^2
    =
    \frac{{\rm Var} (H)}{L^2}.
\end{equation}
Substituting Eq.~\eqref{eq:hamming_variance_exact_sum} and using
\(M=L/b\), we find
\begin{equation}
        \sigma_H^2
        =
        \frac{1}{4L M}
        \left[
            b+2\sum_{r=1}^{b-1}(b-r)a^r
        \right],
        \qquad M=\frac{L}{b}.
    \label{eq:normalized_hamming_variance_exact}
\end{equation}
Equivalently, since \(L=bM\),
\begin{equation}
        \sigma_H^2
        =
        \frac{1}{4b^2M}
        \left[
            b+2\sum_{r=1}^{b-1}(b-r)a^r
        \right].
    \label{eq:normalized_hamming_variance_blocks}
\end{equation}

For fixed block size \(b\) and temperature, the quantity in square
brackets is independent of \(M\). Therefore, the variance of the
normalized Hamming distance is inversely proportional to the number
of blocks:
\begin{equation}
    \sigma_H^2
    \propto \frac{1}{M}.
\end{equation}
Taking the inverse yields
\begin{equation}
        \frac{1}{\sigma_H^2}
        =
        \frac{4b^2}{
            b+2\displaystyle\sum_{r=1}^{b-1}(b-r)a^r
        }
        \frac{L}{b}.
    \label{eq:inverse_normalized_hamming_variance}
\end{equation}
Thus, at fixed \(b\) and temperature, the inverse variance of the
normalized Hamming distance is exactly linear in the number of
independent blocks \(L/b\). The proportionality coefficient depends
on the block size and the interaction strength through
\(a=\tanh^2(\beta J)\).

In the low-temperature limit, \(\beta J\to\infty\), and hence
\(a\to1\). Each block becomes fully magnetized, with its two
possible ground states corresponding to all spins up or all spins
down.

It follows that

\begin{equation}
        \frac{1}{4\sigma_H^2}
        =
        M
        =
        \frac{L}{b}.
    \label{eq:inverse_normalized_hamming_variance_zeroT}
\end{equation}

This establishes that, in the zero-temperature limit, the inverse
variance of the normalized Hamming-distance distribution is
exactly proportional to the number of independent blocks, with
slope \(4\). At finite temperature, the same linear dependence on
\(L/b\) holds when \(b\) and temperature are fixed, but the slope
is modified by the intrablock spin correlations.

\subsection{Distribution of distances on the hyper-sphere}

We consider two independent points \(\mathbf{x}\) and \(\mathbf{y}\)
uniformly distributed on the unit hypersphere
\begin{equation}
    S^{D-1}
    =
    \left\{
        \mathbf{x}\in\mathbb{R}^{D}
        \,:\,
        \|\mathbf{x}\|=1
    \right\}.
\end{equation}
Our goal is to derive the probability density function (PDF) of
the geodesic distance between these points and establish its
relationship with the dimension \(D\) of the embedding space.

The geodesic distance between two points on the unit hypersphere is
the central angle \(\theta\) between their position vectors:
\begin{equation}
    \theta
    =
    \arccos(\mathbf{x}\cdot\mathbf{y}),
    \qquad
    0\leq\theta\leq\pi.
    \label{eq:geodesic_definition}
\end{equation}
Because the uniform distribution on the sphere is invariant under
rotations, we can fix the first point at the north pole without
loss of generality. The distribution of \(\theta\) is then equivalent
to the distribution of the polar angle of a uniformly sampled point
\(\mathbf{y}\) relative to that pole.

To derive this distribution, we use hyperspherical coordinates. The
surface element on \(S^{D-1}\) can be written as
\begin{equation}
    d\Omega_{D-1}
    =
    (\sin\theta)^{D-2}
    \,d\theta\,d\Omega_{D-2},
    \label{eq:hyperspherical_surface_element}
\end{equation}
where \(d\Omega_{D-2}\) denotes the surface element associated with
the remaining angular coordinates. The factor
\((\sin\theta)^{D-2}\) accounts for the surface area of the
hyperspherical slice at polar angle \(\theta\).

Since \(\mathbf{y}\) is uniformly distributed on \(S^{D-1}\), its
probability measure is proportional to the surface element. Integrating
over all angular coordinates other than \(\theta\), we obtain
\begin{equation}
    p_D(\theta)\,d\theta
    =
    \frac{
        \Omega_{D-2}(\sin\theta)^{D-2}\,d\theta
    }{
        \Omega_{D-1}
    },
    \label{eq:geodesic_probability_element}
\end{equation}
where \(\Omega_{D-1}\) is the total surface area of the unit sphere
\(S^{D-1}\). The surface area of a unit sphere of dimension \(n\) is
\begin{equation}
    \Omega_n
    =
    \frac{2\pi^{(n+1)/2}}
    {\Gamma\left(\frac{n+1}{2}\right)}.
    \label{eq:hypersphere_surface_area}
\end{equation}
Consequently,
\begin{equation}
    \frac{\Omega_{D-2}}{\Omega_{D-1}}
    =
    \frac{
        \Gamma\left(\frac{D}{2}\right)
    }{
        \sqrt{\pi}\,
        \Gamma\left(\frac{D-1}{2}\right)
    }.
\end{equation}

The resulting probability density of the geodesic distance is
therefore
\begin{equation}
        p_D(\theta)
        =
        \frac{
            \Gamma\left(\frac{D}{2}\right)
        }{
            \sqrt{\pi}\,
            \Gamma\left(\frac{D-1}{2}\right)
        }
        (\sin\theta)^{D-2},
        \qquad 0\leq\theta\leq\pi.
    \label{eq:geodesic_pdf}
\end{equation}
The normalization follows directly from the ratio of sphere surface
areas, or equivalently from
\begin{equation}
    \int_0^\pi(\sin\theta)^{D-2}\,d\theta
    =
    \frac{
        \sqrt{\pi}\,
        \Gamma\left(\frac{D-1}{2}\right)
    }{
        \Gamma\left(\frac{D}{2}\right)
    }.
\end{equation}

An equivalent derivation can be obtained by considering the scalar
product \(u=\mathbf{x}\cdot\mathbf{y}=\cos\theta\). Its PDF is
\begin{equation}
    f_D(u)
    =
    \frac{
        \Gamma\left(\frac{D}{2}\right)
    }{
        \sqrt{\pi}\,
        \Gamma\left(\frac{D-1}{2}\right)
    }
    (1-u^2)^{(D-3)/2},
    \qquad -1\leq u\leq 1.
    \label{eq:dot_product_pdf}
\end{equation}
Applying the change of variables \(u=\cos\theta\), for which
\(\lvert du/d\theta\rvert=\sin\theta\), recovers
Eq.~\eqref{eq:geodesic_pdf}.

The \(n\)-th moment of the geodesic distance is defined as
\begin{equation}
    \langle\theta^n\rangle
    =
    \int_0^\pi \theta^n p_D(\theta)\,d\theta.
    \label{eq:geodesic_moment}
\end{equation}
In particular, the mean is
\begin{equation}
    \langle\theta\rangle
    =
    \int_0^\pi \theta p_D(\theta)\,d\theta.
\end{equation}

The distribution is symmetric about \(\theta=\pi/2\), since
\begin{equation}
    p_D(\pi-\theta)=p_D(\theta).
    \label{eq:geodesic_symmetry}
\end{equation}
Using this symmetry, we find
\begin{align}
    \langle\theta\rangle
    &=
    \int_0^{\pi/2}
    \left[
        \theta p_D(\theta)
        +(\pi-\theta)p_D(\pi-\theta)
    \right]d\theta
    \nonumber\\
    &=
    \pi\int_0^{\pi/2}p_D(\theta)\,d\theta
    =
    \frac{\pi}{2}.
\end{align}
Thus,
\begin{equation}
        \langle\theta\rangle=\frac{\pi}{2}
    \label{eq:geodesic_mean}
\end{equation}
for every \(D\geq2\). The mean geodesic distance is independent
of the embedding dimension.

The variance is defined as
\begin{equation}
    \operatorname{Var}(\theta)
    =
    \langle\theta^2\rangle-\langle\theta\rangle^2.
\end{equation}
Using the mean obtained above, this becomes
\begin{equation}
    \operatorname{Var}(\theta)
    =
    \left\langle
        \left(\theta-\frac{\pi}{2}\right)^2
    \right\rangle.
    \label{eq:geodesic_variance}
\end{equation}

Introducing \(z=\theta-\pi/2\), we obtain
\begin{equation}
    \operatorname{Var}(\theta)
    =
    C_D
    \int_{-\pi/2}^{\pi/2}
    z^2(\cos z)^{D-2}\,dz,
    \label{eq:geodesic_variance_integral}
\end{equation}
where
\begin{equation}
    C_D
    =
    \frac{
        \Gamma\left(\frac{D}{2}\right)
    }{
        \sqrt{\pi}\,
        \Gamma\left(\frac{D-1}{2}\right)
    }.
\end{equation}
Although this integral does not generally yield a simple elementary
expression, its dependence on \(D\) can be established in the
large-dimensional limit.

For large \(D\), the factor \((\sin\theta)^{D-2}\) strongly
suppresses values of \(\theta\) away from \(\pi/2\). To quantify this
concentration, we expand the distribution around its maximum by
introducing \(z=\theta-\pi/2\). Since
\begin{equation}
    \sin\theta=\cos z,
\end{equation}
the Taylor expansion of the logarithm gives
\begin{equation}
    \log(\cos z)
    =
    -\frac{z^2}{2}
    -\frac{z^4}{12}
    +\mathcal{O}(z^6).
    \label{eq:log_cos_expansion}
\end{equation}
It follows that
\begin{equation}
    (\cos z)^{D-2}
    =
    \exp\left[
        -\frac{D-2}{2}z^2
        -\frac{D-2}{12}z^4
        +\mathcal{O}(Dz^6)
    \right].
    \label{eq:cos_power_expansion}
\end{equation}

The dominant contribution comes from \(z=\mathcal{O}(D^{-1/2})\).
To leading order, the distribution of \(z\) is therefore Gaussian:
\begin{equation}
    p_D(z)
    \simeq
    \sqrt{\frac{D}{2\pi}}
    \exp\left(-\frac{Dz^2}{2}\right).
    \label{eq:geodesic_gaussian_approximation}
\end{equation}
Since a Gaussian distribution proportional to
\(\exp(-Dz^2/2)\) has variance \(1/D\), we conclude that
\begin{equation}
        \operatorname{Var}(\theta)
        =
        \frac{1}{D}
        +\mathcal{O}\left(\frac{1}{D^2}\right).
    \label{eq:geodesic_variance_asymptotic}
\end{equation}

The results show that the mean geodesic distance remains fixed at
\(\pi/2\), while its variance decreases inversely with the embedding
dimension. In high-dimensional spaces, two independently sampled
points are therefore typically separated by an angle close to
\(\pi/2\), with fluctuations that vanish as \(D^{-1/2}\).

The inverse-dimensional scaling provides an asymptotic relationship
between the variance of pairwise geodesic distances and the embedding
dimension:
\begin{equation}
        D
        \simeq
        \frac{1}{\operatorname{Var}(\theta)}.
    \label{eq:dimension_estimator}
\end{equation}
This relation can be used to estimate the embedding dimension from
the empirical variance of geodesic distances when the points are
approximately independent and uniformly distributed on the sphere.

\subsection{Lineage-Through-Time}
\begin{figure*}[h!]
        \centering
        \includegraphics[width=7in]{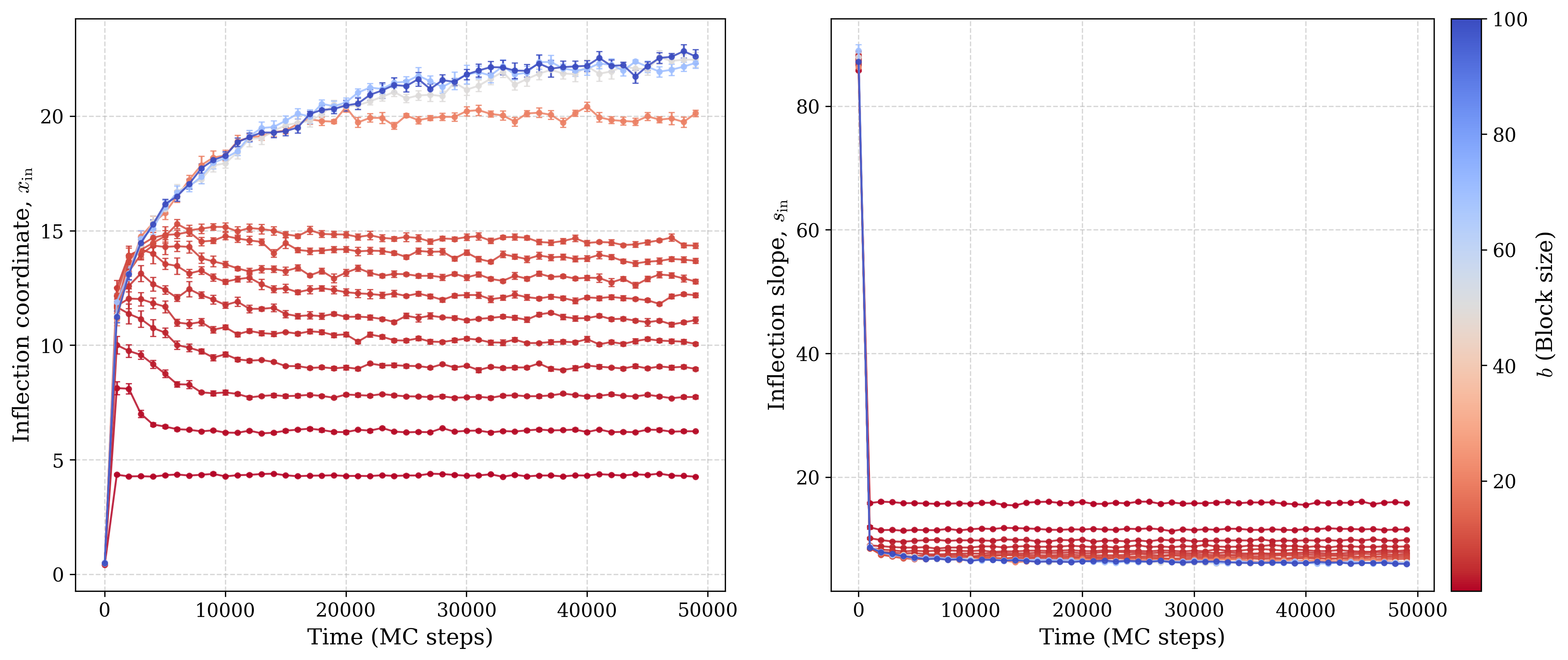}
       \caption{Trajectories of the lineage-through-time (LTT) feature of the star phylogenies as a function of the MC time. Left: value of the inflection point horizontal coordinate ($x_{\rm in}$). Right: value of the LTT curve slope at the inflection point ($s_{\rm in}$). Colors indicate different values of the block size $b$, as shown in the colorbar. The limit values correspond to the one reported in the main text.}
        \label{fig:SI_ising_dIN_vs_time}
\end{figure*}


\newpage

\subsection{$d_{\rm IN}$ trajectories along a star-phylogeny}
\begin{figure*}[h!]
        \centering
        \includegraphics[width=7in]{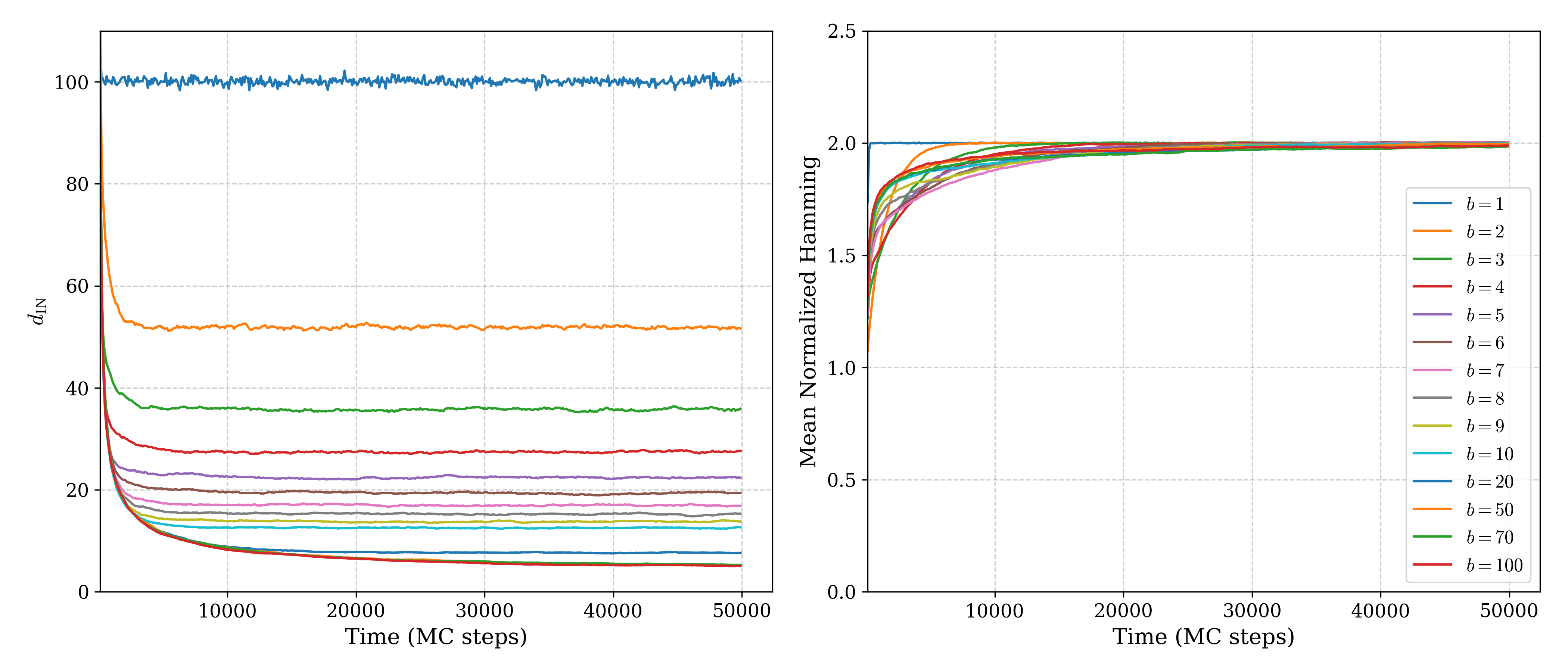}
       \caption{(Left): Trajectories of the intrinsic dimension of the star phylogenies, computed using the variance of the normalized Hamming distance via Eq. (2) in the main text for the 100 independent walkers as a function of time. The limit value agrees with the canonical limit reported in the paper. (Right): Average of the normalized Hamming distance; the limit is independent of $b$ and approaches the theoretical value of 2.}
        \label{fig:SI_ising_dIN_vs_time}
\end{figure*}

\subsection{Partition function and recurrence time}

We can analytically compute the partition function of our toy model by using the standard results for the 1D Ising model, as follows.

Consider a one-dimensional spin system of length \(L\), with
\(s_i\in\{-1,+1\}\), whose Hamiltonian is block structured. Let
\(b\) be the block size and assume that \(b\) divides \(L\), so
that the system consists of
\begin{equation}
    M=\frac{L}{b}
\end{equation}
blocks, each containing \(b\) spins.

Spins within the same block interact through nearest-neighbor
ferromagnetic couplings, while spins belonging to different blocks
do not interact. The Hamiltonian can therefore be written as
\begin{equation}
    \mathcal{H}(\mathbf{s})
    =
    -J\sum_{m=0}^{M-1}
    \sum_{j=1}^{b-1}
    s_{mb+j}s_{mb+j+1},
    \label{eq:block_ising_hamiltonian}
\end{equation}
where \(J>0\) is the ferromagnetic coupling constant. We initially
assume open boundary conditions within each block.

The canonical partition function is
\begin{equation}
    Z_L
    =
    \sum_{\{s_i=\pm1\}}
    e^{-\beta\mathcal{H}(\mathbf{s})},
    \qquad
    \beta=\frac{1}{k_{\mathrm B}T}.
    \label{eq:block_partition_definition}
\end{equation}

Since there are no interactions between different blocks, the
Hamiltonian is a sum of independent block contributions:
\begin{equation}
    \mathcal{H}(\mathbf{s})
    =
    \sum_{m=0}^{M-1}\mathcal{H}_m,
\end{equation}
where
\begin{equation}
    \mathcal{H}_m
    =
    -J\sum_{j=1}^{b-1}
    s_{mb+j}s_{mb+j+1}.
\end{equation}

Consequently, the Boltzmann factor factorizes:
\begin{equation}
    e^{-\beta\mathcal{H}(\mathbf{s})}
    =
    \prod_{m=0}^{M-1}e^{-\beta\mathcal{H}_m}.
\end{equation}
Because the spin sums over different blocks are independent, the
partition function becomes
\begin{align}
    Z_L
    &=
    \prod_{m=0}^{M-1}
    \left[
        \sum_{\{s_{mb+j}=\pm1\}_{j=1}^{b}}
        e^{-\beta\mathcal{H}_m}
    \right]
    \nonumber\\
    &=
    \left(Z_b\right)^M,
    \label{eq:block_partition_factorization}
\end{align}
where \(Z_b\) is the partition function of a
single open Ising chain of length \(b\).

For a single block, the Hamiltonian is
\begin{equation}
    \mathcal{H}_b
    =
    -J\sum_{j=1}^{b-1}s_js_{j+1}.
\end{equation}
Its partition function is
\begin{equation}
    Z_b
    =
    \sum_{\{s_j=\pm1\}}
    \exp\left(
        \beta J\sum_{j=1}^{b-1}s_js_{j+1}
    \right).
    \label{eq:single_block_partition}
\end{equation}

Introduce the bond variables
\begin{equation}
    \tau_j=s_js_{j+1},
    \qquad j=1,\ldots,b-1.
\end{equation}
Each bond variable takes values \(\tau_j=\pm1\). For any
assignment of the \(b-1\) bond variables, there are exactly two
compatible spin configurations, corresponding to the two possible
values of the first spin \(s_1\). Thus, the sum over the \(2^b\)
spin configurations can be replaced by a sum over the \(2^{b-1}\)
bond configurations, multiplied by a factor of two:
\begin{align}
    Z_b
    &=
    2\sum_{\{\tau_j=\pm1\}}
    \exp\left(
        \beta J\sum_{j=1}^{b-1}\tau_j
    \right)
    \nonumber\\
    &=
    2\prod_{j=1}^{b-1}
    \left(
        \sum_{\tau_j=\pm1}e^{\beta J\tau_j}
    \right)
    \nonumber\\
    &=
    2\left(e^{\beta J}+e^{-\beta J}\right)^{b-1}.
\end{align}
Using the definition of the hyperbolic cosine, we obtain
\begin{equation}
        Z_b
        =
        2\left[2\cosh(\beta J)\right]^{b-1}.
    \label{eq:open_block_partition_exact}
\end{equation}

Substituting Eq.~\eqref{eq:open_block_partition_exact} into
Eq.~\eqref{eq:block_partition_factorization}, we obtain the exact
partition function of the full system:
\begin{align}
    Z_L
    &=
    \left\{
        2\left[2\cosh(\beta J)\right]^{b-1}
    \right\}^{L/b}.
\end{align}
Equivalently,
\begin{equation}
        Z_L
        =
        2^{L/b}
        \left[2\cosh(\beta J)\right]^{L-L/b}
    \label{eq:block_partition_exact}
\end{equation}

This expression is exact for open boundary conditions within each
block, with no interactions between blocks.

The corresponding Helmholtz free energy is
\begin{equation}
    F_L
    =
    -k_{\mathrm B}T\ln Z_L,
\end{equation}
which gives
\begin{equation}
        F_L
        =
        -k_{\mathrm B}T
        \left[
            \frac{L}{b}\ln 2
            +
            \left(L-\frac{L}{b}\right)
            \ln\left(2\cosh(\beta J)\right)
        \right].
    \label{eq:block_free_energy}
\end{equation}

In the low-temperature limit, the hyperbolic cosine satisfies
\begin{equation}
    2\cosh(\beta J)
    =
    e^{\beta J}+e^{-\beta J}
    \simeq e^{\beta J},
    \qquad \beta J\gg 1.
\end{equation}
Therefore,
\begin{equation}
    \ln\left[2\cosh(\beta J)\right]
    \simeq \beta J,
\end{equation}
and the logarithm of the partition function becomes
\begin{align}
    \ln Z_L
    &\simeq
    \frac{L}{b}\ln 2
    +
    \beta J\left(L-\frac{L}{b}\right)
    \nonumber\\
    &=
    \beta JL
    +
    \frac{L}{b}\left(\ln 2-\beta J\right).
    \label{eq:log_partition_lowT}
\end{align}

At fixed total system size \(L\) and temperature \(T\), this expression
is linear in the number of blocks \(M=L/b\):
\begin{equation}
        \ln Z_L
        \simeq
        \beta JL
        +
        M\left(\ln 2-\beta J\right).
    \label{eq:log_partition_linear_blocks}
\end{equation}
The first term is independent of the block size, whereas the second
term describes the dependence on the number of independent blocks.
Equivalently, the partition function takes the asymptotic form
\begin{equation}
        Z_L
        \simeq
        e^{\beta JL}
        \exp\left[
            \frac{L}{b}\left(\ln 2-\beta J\right)
        \right].
    \label{eq:partition_exponential_blocks}
\end{equation}

This result has a direct physical interpretation. At zero temperature,
each block has two degenerate ground states, corresponding to all
spins aligned upward or downward. The ground-state energy of the
full system is
\begin{equation}
    E_0=-J\left(L-\frac{L}{b}\right),
\end{equation}
and the ground-state degeneracy is
\begin{equation}
    g_0=2^{L/b}.
\end{equation}
Consequently, in the zero-temperature limit, the partition function
is dominated by the ground states:
\begin{equation}
    Z_L
    \sim
    g_0 e^{-\beta E_0}
    =
    2^{L/b}
    \exp\left[
        \beta J\left(L-\frac{L}{b}\right)
    \right].
\end{equation}

To isolate the contribution of the ground-state degeneracy, we may
remove the ground-state Boltzmann factor and define the normalized
partition function
\begin{equation}
    \widetilde{Z}_L
    =
    Z_L e^{\beta E_0}.
\end{equation}
In the zero-temperature limit,
\begin{equation}
    \widetilde{Z}_L
    \longrightarrow
    2^{L/b},
\end{equation}
so that
\begin{equation}
        \ln \widetilde{Z}_L
        \longrightarrow
        \frac{L}{b}\ln 2.
    \label{eq:normalized_log_partition_lowT}
\end{equation}

Thus, the logarithm of the normalized partition function varies
exactly linearly with the number of blocks in the low-temperature
limit, with slope \(\ln 2\). For the full partition function, the
dependence on \(L/b\) is also linear at low temperature, but its
slope is \(\ln 2-\beta J\) when \(L\) is held fixed.

Assuming that the estimated \(d_{\rm IN}\) is also valid locally around typical sequences, the number of available states within a radius \(R\) of a given sequence can be approximated by the volume of the corresponding \(d_{\rm IN}\)-dimensional ball, \(A\). Accordingly, the partition function restricted to this ball scales as
\begin{equation}
    Z_A \propto R^{d_{\rm IN}}.
    \label{eq:local_partition}
\end{equation}

Combining Eqs.~\eqref{eq:local_partition} and \eqref{eq:normalized_log_partition_lowT}, we obtain an estimate of the local Boltzmann measure, \(\pi(A)\), of the set \(A\) around a typical sequence:
\begin{equation}
    \pi(A) = \frac{Z_A}{\widetilde{Z}_L}
    \sim \frac{R^{d_{\rm IN}}}{2^{L/b}}.
    \label{eq:local_measure}
\end{equation}

Invoking Kac's lemma and using the scaling \(d_{\rm IN}\sim L/b\), we can estimate the recurrence time of the set \(A\) as
\begin{equation}
    \tau_{\rm rec} \propto \frac{1}{\pi(A)}
    \sim \left(\frac{2}{R}\right)^{d_{\rm IN}}.
    \label{eq:local_recurrence}
\end{equation}

\begin{figure*}[h]
        \centering
        \includegraphics[width=4in]{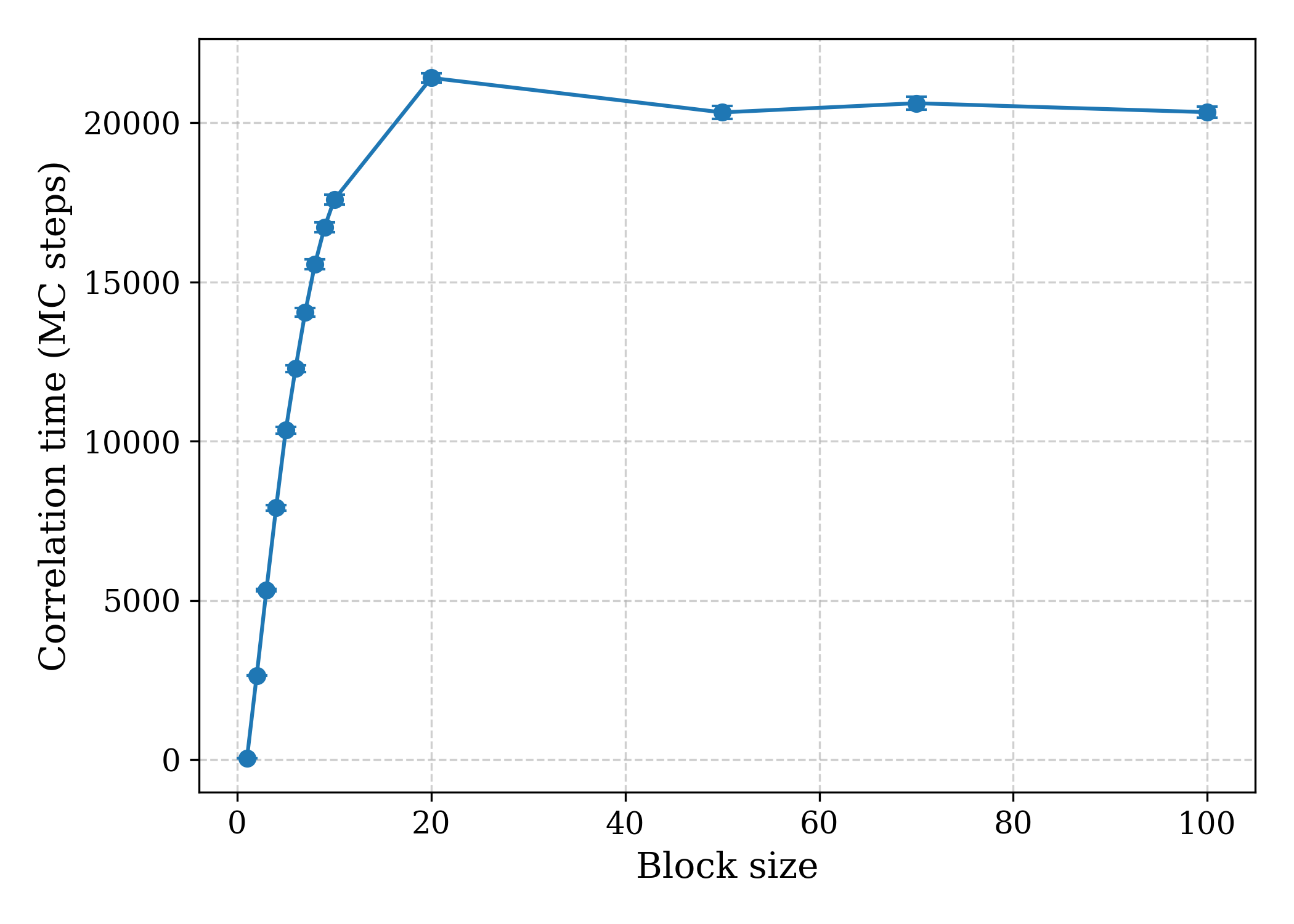}
       \caption{Correlation time of the Markov chain as a function of the block size. Larger blocks make the energy contribution more significant and slow down the chain dynamics.}
        \label{fig:SI_ising_taucorr}
\end{figure*}

Taking the autocorrelation time, \(\tau_{\rm corr}\), as the characteristic timescale of the Markov chain (see Fig.~\ref{fig:SI_ising_taucorr}), we can express the recurrence-time scaling more generally as
\begin{equation}
    \log\left(\frac{\tau_{\rm rec}}{\tau_{\rm corr}}\right)
    \sim \kappa(R)*d_{\rm IN}+\mathrm{const}.
    \label{eq:recurrence_scaling}
\end{equation}

\newpage

\section{Potts model for natural proteins}
\subsection{Simulation details}
We fitted the Potts model starting from the PFAM family PF00520 sequences. We first computed the natural MSA of the family and filtered it using a normalized Hamming distance threshold of 0.2 and removing sequences with a gap fraction higher than 80\%. This procedure is intended to remove spurious phylogenetic signal in the natural sequences and clean the dataset from noise.

Using the 45000 remaining sequences in the filtered MSA, we used the Mi3 software (ref.~[38]) to compute the Potts couplings associated with the observed single- and pairwise frequencies, resulting in a one-hot-encoded Hamiltonian of length $L=188$ and alphabet size $q=21$ (20 amino acids + 1 gap symbol).

Once we obtained the Potts Hamiltonian, we performed star-phylogeny Monte Carlo simulations starting from 10 uniformly sampled initial sequences and simulating 100 independent walkers from each initial condition, for a total of 1000 independent walkers. Each walker was simulated for $10^7$ timesteps.

The second half of these trajectories was sampled every 100000 steps (longer than the chain autocorrelation time; see Fig.~\ref{fig:SI_potts_tau_corr}) to obtain 50000 uncorrelated sequences, comparable to the size of the available MSA.
The resulting canonical sample was used to compute the Hamming distance distribution and the intrinsic dimension of the dataset.

Full star-phylogeny trajectories were used to estimate the LTT features as well as the recurrence probability of the chain.

This procedure was repeated for each value of the shuffled columns.


\subsection{Hamming distance}
\begin{figure*}[h]
       \centering
       \includegraphics[width=7in]{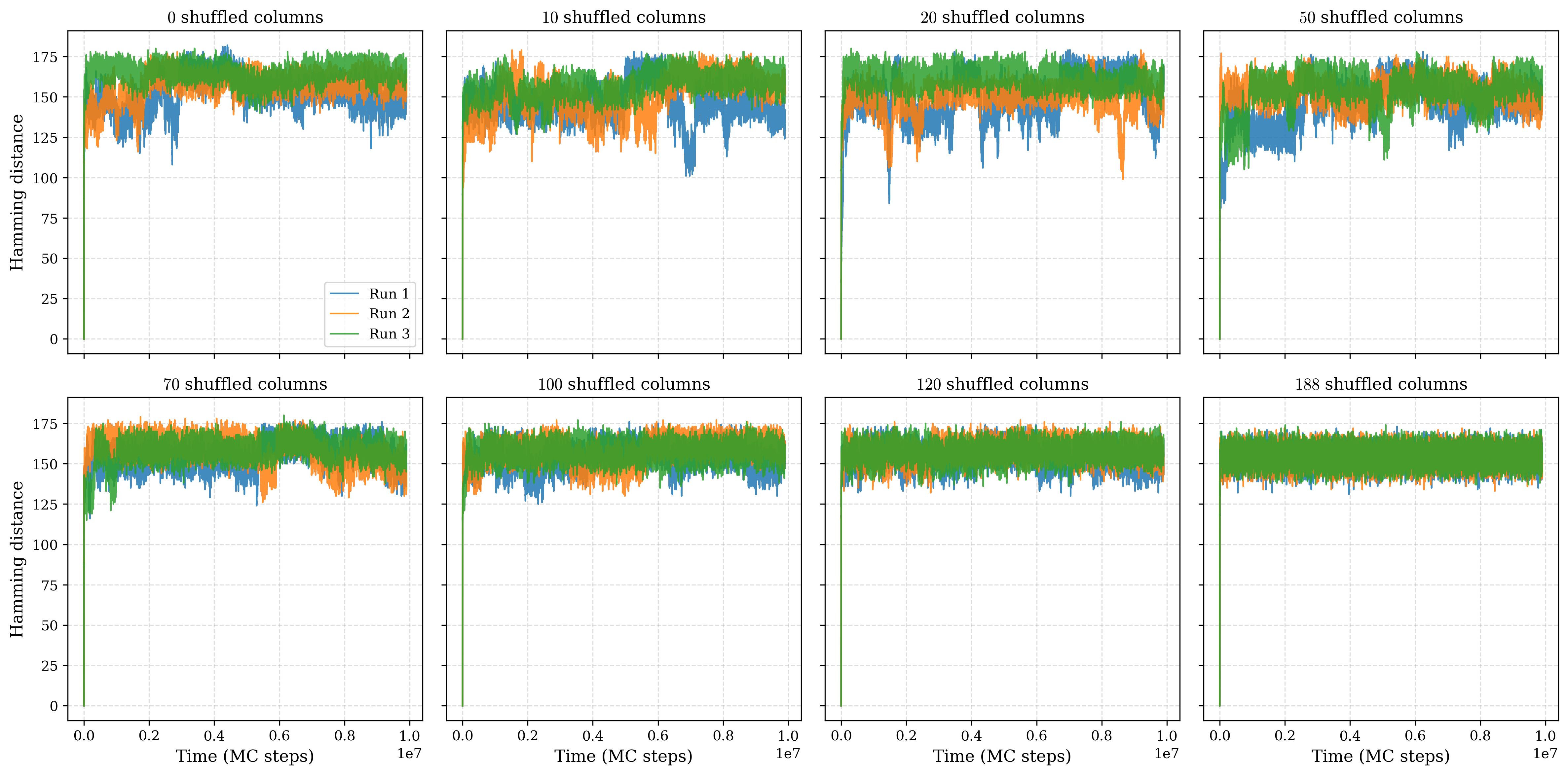}
      \caption{Trajectories of Hamming distances from the initial sequence for different numbers of shuffled columns.}
       \label{fig:SI_potts_hamming_traj}
\end{figure*}

\begin{figure*}[h]
       \centering
       \includegraphics[width=4in]{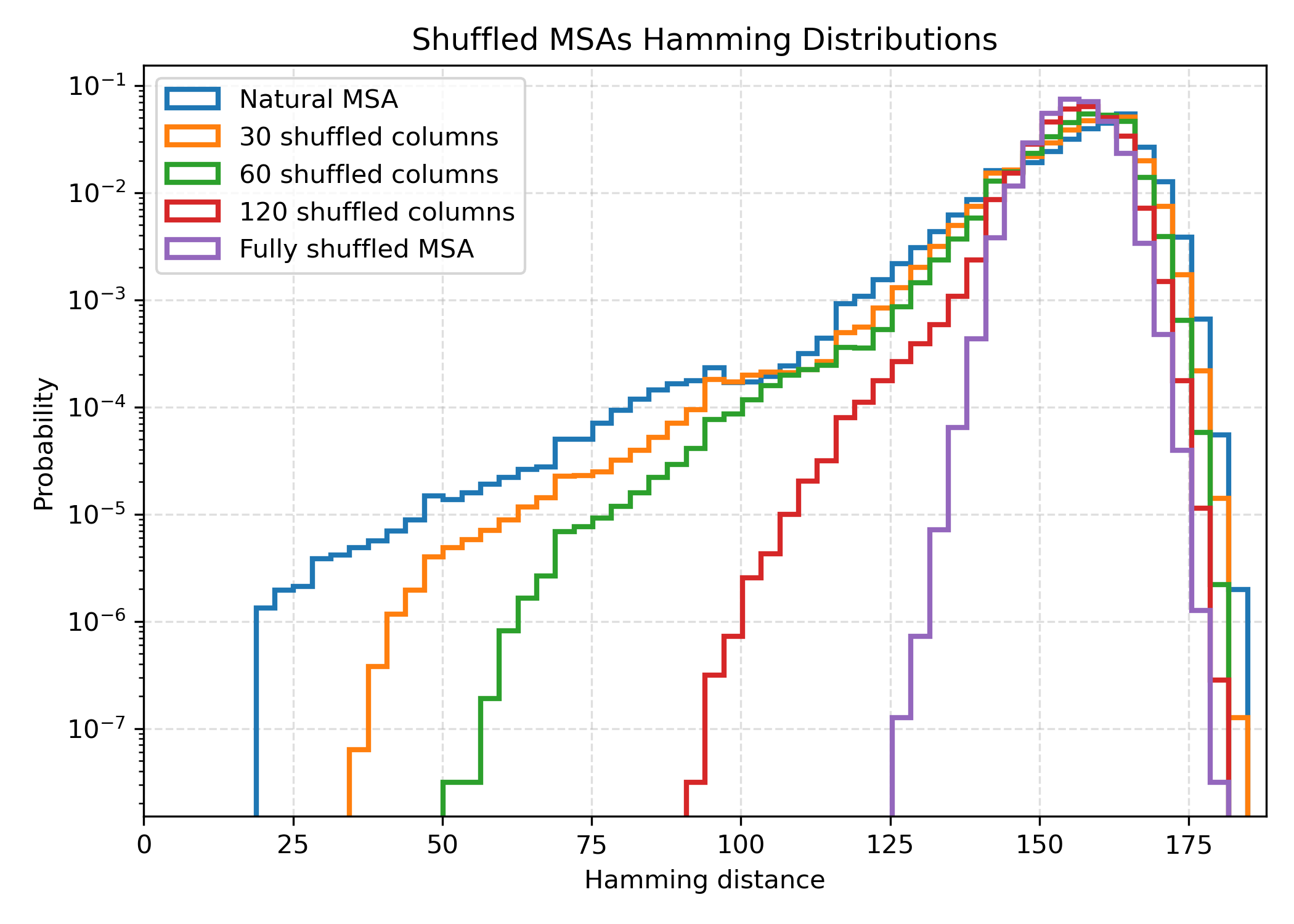}
      \caption{Hamming distance distributions for different degrees of shuffling of the natural MSA.}
       \label{fig:SI_potts_histogram_shuffled}
\end{figure*}

\begin{figure*}[h]
       \centering
       \includegraphics[width=4in]{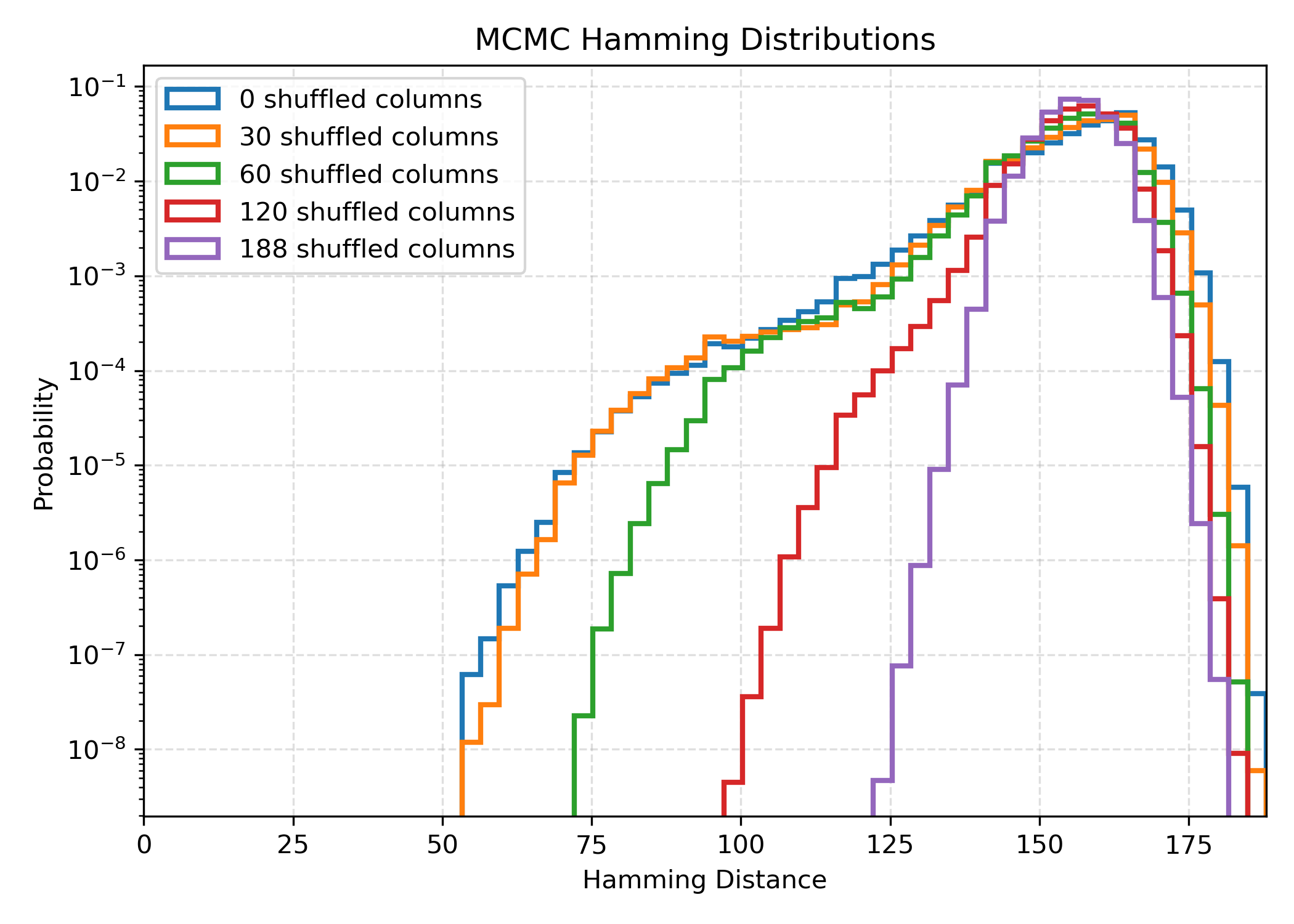}
      \caption{Hamming distance distributions for different MCMC Potts simulations fitted using varying degrees of shuffling of the natural MSA. Small distances present in the MSA distributions were cut before fitting the Potts Hamiltonians because they were interpreted as phylogenetic signal in the MSA and are not reproduced in the simulations.}
       \label{fig:SI_potts_MC_histogram_shuffled}
\end{figure*}

\begin{figure*}[h!]
        \centering
        \includegraphics[width=7in]{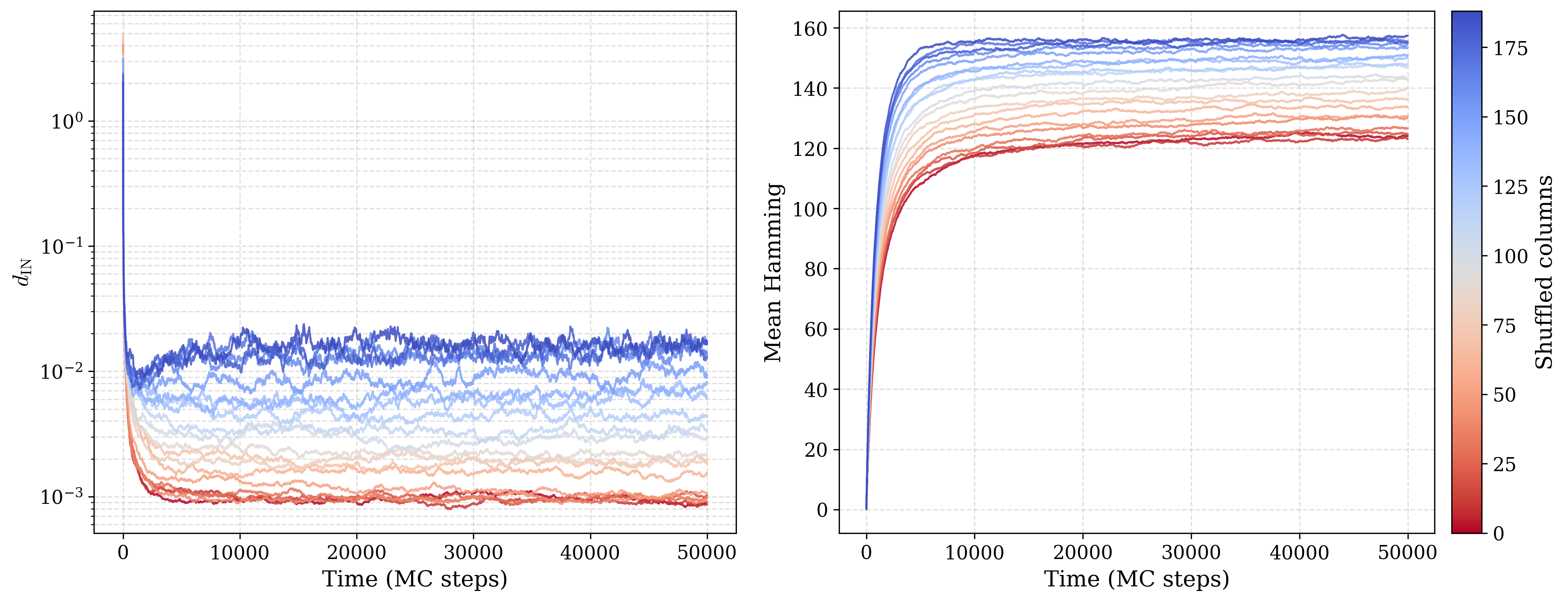}
       \caption{(Left): Trajectories of the intrinsic dimension of the star phylogenies, computed using the variance of the normalized Hamming distance via Eq. (2) in the main text for the 100 independent walkers as a function of time. The limit value are comparable with the canonical limit reported in the paper. (Right): Average of the normalized Hamming distance; the limit is independent of $b$ and approaches the theoretical value of 2.}
        \label{fig:SI_ising_dIN_vs_time}
\end{figure*}


\subsection{Frobenius Norm of Covariance Matrix in the Fully Shuffled Case}

In the fully shuffled limit, the MSA is described by an independent-site model, in which the symbols at different positions are statistically independent while the single-site frequencies are preserved. Let \(p_i(a)\) denote the frequency of state \(a\) at position \(i\), with \(a=1,\ldots,q\), and let \(x_i^a\) be the corresponding one-hot encoded variable. The covariance matrix is defined as
\begin{equation}
    C_{ij}^{ab}
    =
    \langle x_i^a x_j^b\rangle
    -
    \langle x_i^a\rangle\langle x_j^b\rangle.
\end{equation}
In the independent-site limit, the covariance between distinct positions vanishes, \(C_{ij}^{ab}=0\) for \(i\neq j\), whereas the covariance within a single position is given by
\begin{equation}
    C_{ii}^{ab}
    =
    p_i(a)\delta_{ab}-p_i(a)p_i(b).
\end{equation}
Consequently, the full covariance matrix is block diagonal, and its squared Frobenius norm reduces to the sum of the squared Frobenius norms of the single-site covariance blocks:
\begin{equation}
    \|C\|_F^2
    =
    \sum_{i=1}^{L}\sum_{a,b=1}^{q}
    \left[p_i(a)\delta_{ab}-p_i(a)p_i(b)\right]^2.
\end{equation}
Expanding the square and using \(\sum_a p_i(a)=1\), we obtain
\begin{equation}
    \|C\|_F^2
    =
    \sum_{i=1}^{L}
    \left[
    \sum_{a=1}^{q}p_i(a)^2
    -2\sum_{a=1}^{q}p_i(a)^3
    +\left(\sum_{a=1}^{q}p_i(a)^2\right)^2
    \right].
\end{equation}

This expression provides the expected baseline for the full one-hot covariance norm after complete shuffling. It retains the contribution of single-site fluctuations while eliminating all inter-site covariance contributions.


\subsection{Alternative $d_{\rm IN}$ estimators}
\begin{figure*}[h]
        \centering
        \includegraphics[width=7in]{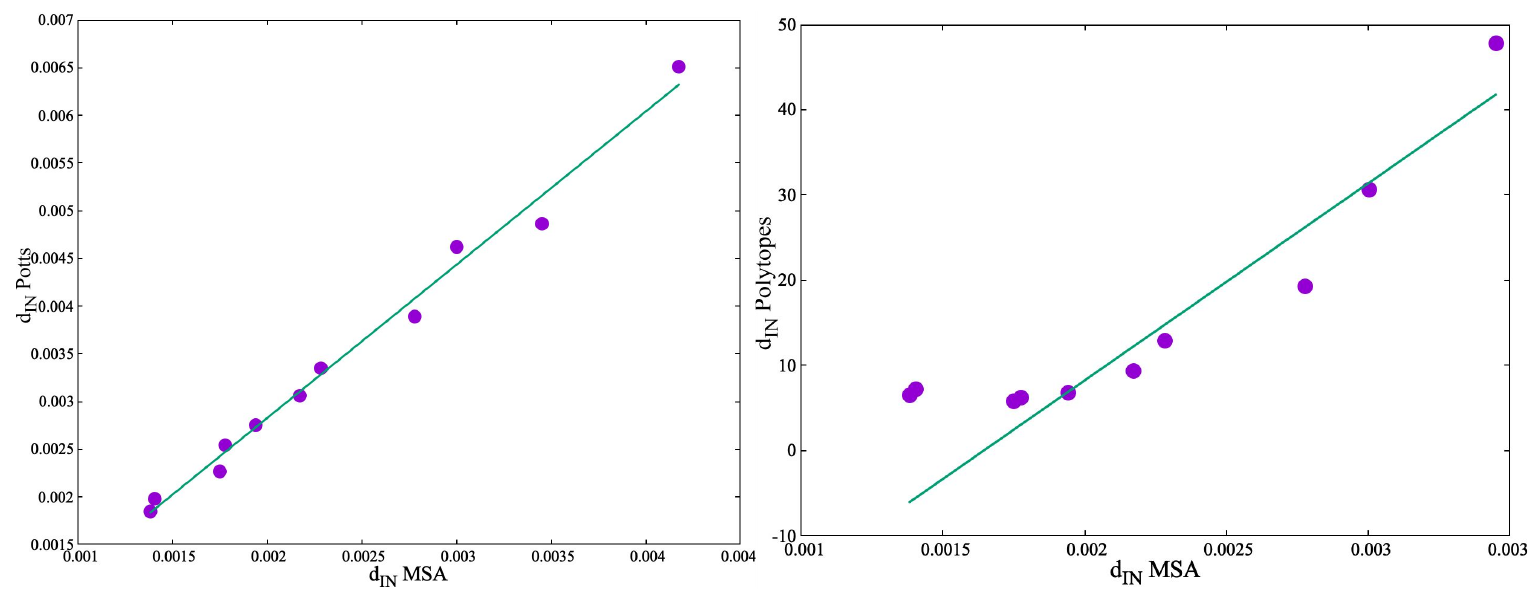}
       \caption{(Left) Scatter plot of $d_{\rm IN}$ estimated from the MSA and from a set of uncorrelated sequences obtained from the Potts simulations. The linear dependence and the comparable scales suggest that the influence of the residual phylogenetic signal in the MSA does not affect the intrinsic dimension estimation significantly. (Right) Scatter plot of $d_{\rm IN}$ estimated from the MSA and the polytope-based method developed in Ref.~[45]. This last estimator appears to diverge from our estimated dependence and plateau in the low-dimensional region, while showing partial agreement across the remainder of the support. Nevertheless, the overall increasing trend is preserved in all analyzed cases.}
        \label{fig:SI_potts_tau_corr}
\end{figure*}

\newpage

\subsection{Autocorrelation and recurrence time}
\begin{figure*}[h]
        \centering
        \includegraphics[width=7in]{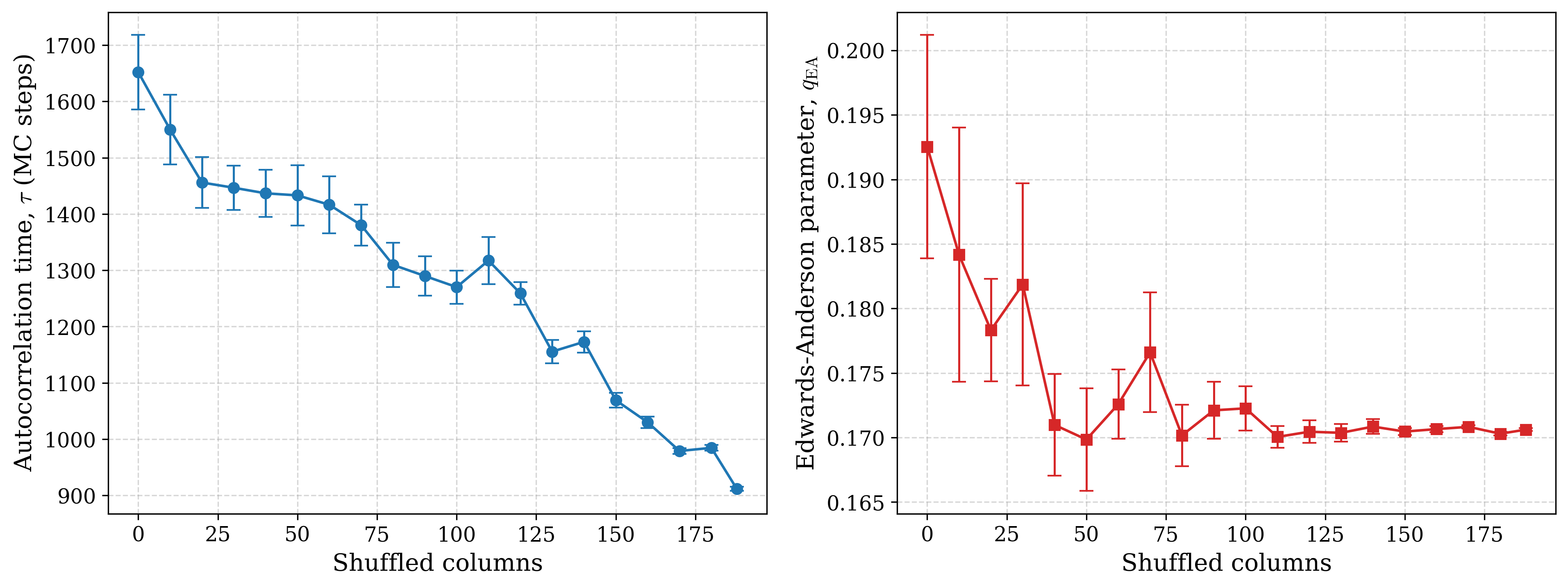}
       \caption{(Left) Autocorrelation time	$\tau_{\rm corr}$ for MC with Glauber dynamics as a function of the number of shuffled columns in the MSA. This time was measured by shifting and rescaling the correlation functions and quantifies the intrabasin relaxation of the system. (Right) Edwards-Anderson parameter as a function of the shuffled columns in the MSA, measured as the asymptotic value of the unshifted autocorrelation functions. The presence of strongly conserved positions renders the asymptotic autocorrelation of the dynamics non-zero.}
        \label{fig:SI_potts_tau_corr}
\end{figure*}

\begin{figure*}[h]
        \centering
        \includegraphics[width=4in]{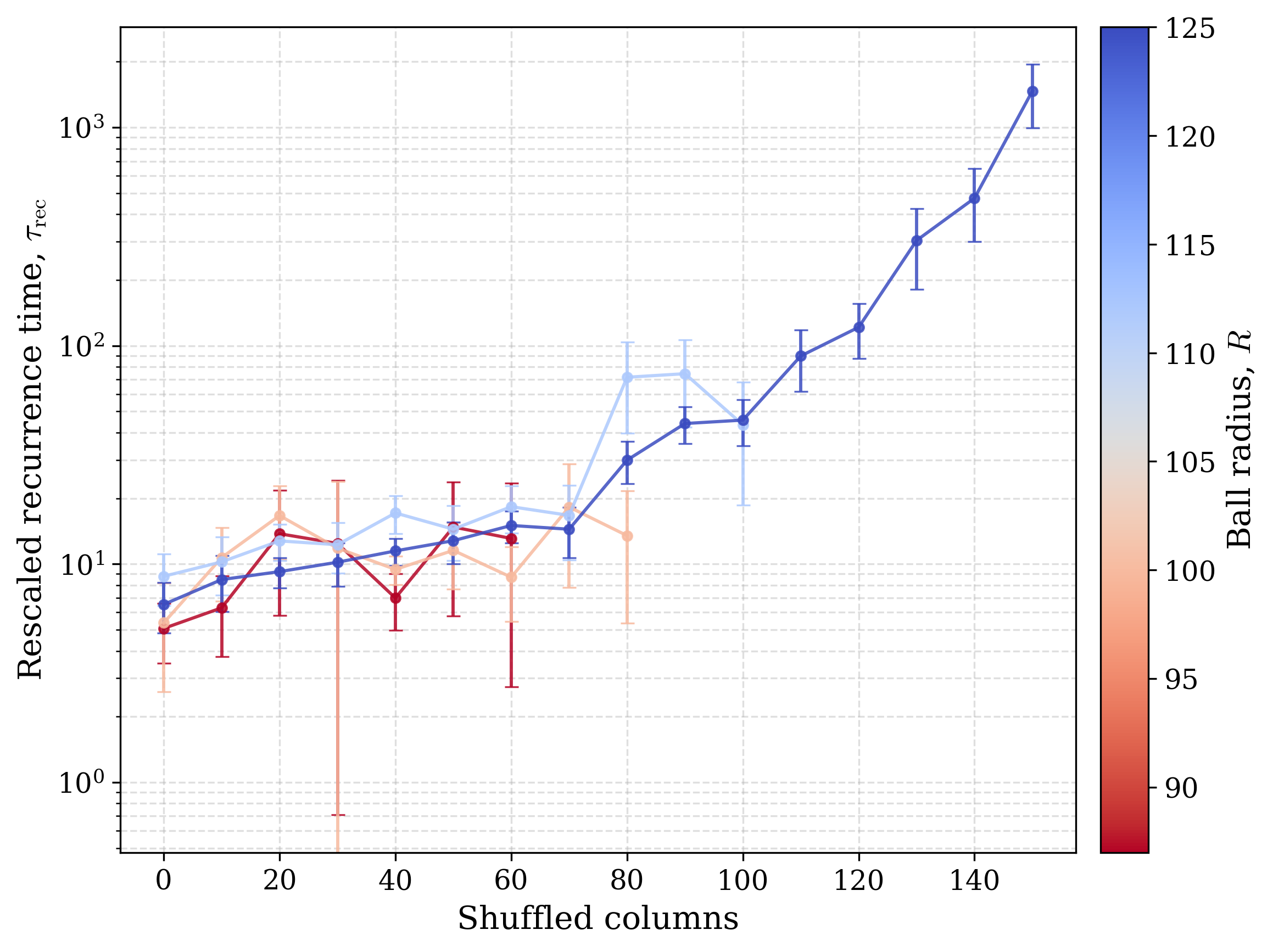}
       \caption{Rescaled recurrence time as a function of the shuffled columns in the MSA for different values of the recurrence radius $R$. The large nominal sequence space ($L=188, q=21$) and the fast decaying Hamming distributions (see Fig.~\ref{fig:SI_potts_histogram_shuffled}) make the estimation of the recurrence time computationally hard for small radii. For the explored values of $R$, the dependence of the recurrence time on the shuffling appears robust.}
        \label{fig:SI_potts_tau_rec}
\end{figure*}

\begin{figure*}[h!]
        \centering
        \includegraphics[width=4in]{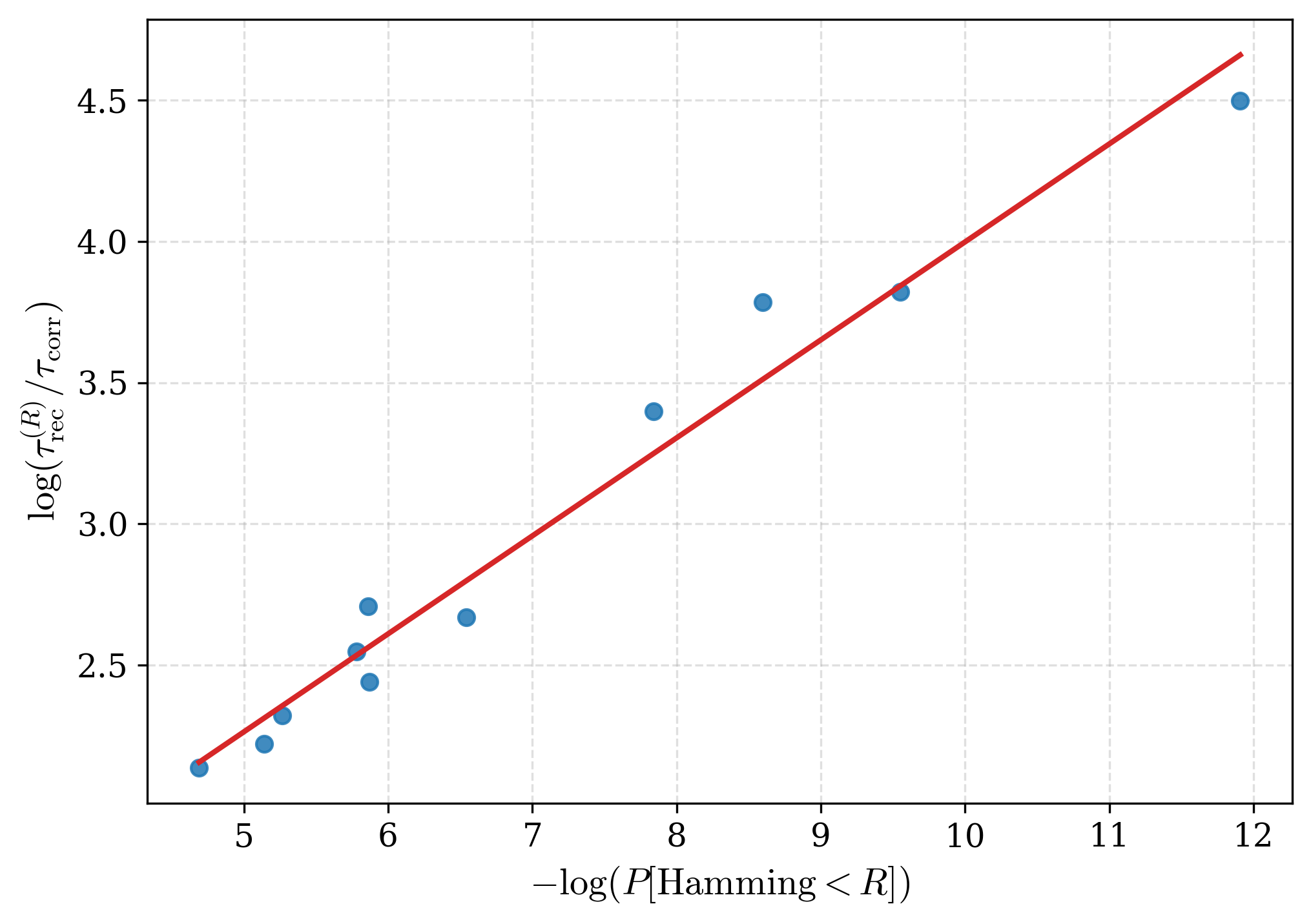}
       \caption{Scatter plot of the logarithm of the left tail below a threshold $R$ of the Hamming distance distribution against the logarithm of the rescaled recurrence time in a ball of the same radius $R$. The linear dependence supports the applicability of the Kac's lemma approximation in the context of the complex landscapes of natural proteins, suggesting that the volume of the ball correctly scales with the intrinsic dimension of the sequence space. The Hamming distance probability was computed using a set of uncorrelated sequences from the MC simulations, thereby accounting for the phylogenetic signal that may be left in the MSA even after the filtering procedure.}
        \label{fig:SI_potts_logprob_vs_logtau}
\end{figure*}

\end{document}